\pdfoutput=1

\documentclass[11pt,twoside,a4paper,cmspaper,final,collab]{cms-tdr}
\def\svnVersion{367093}\def\cmsCernNoTag{CERN-PH-EP/2015-309}\def\cmsCernDate{\today}\def\cmsMessage{Published in the Journal of High Energy Physics as \href{http://dx.doi.org/10.1007/JHEP08(2016)139}{\doi{10.1007/JHEP08(2016)139.}}}

\begin{document}\cmsNoteHeader{FSQ-12-002}

\hyphenation{had-ron-i-za-tion}
\hyphenation{cal-or-i-me-ter}
\hyphenation{de-vices}
\RCS$Revision: 363379 $
\RCS$HeadURL: svn+ssh://svn.cern.ch/reps/tdr2/papers/FSQ-12-002/trunk/FSQ-12-002.tex $
\RCS$Id: FSQ-12-002.tex 363379 2016-08-04 05:42:50Z ivanp $
\newlength\cmsFigWidth
\ifthenelse{\boolean{cms@external}}{\setlength\cmsFigWidth{0.85\columnwidth}}{\setlength\cmsFigWidth{0.4\textwidth}}
\ifthenelse{\boolean{cms@external}}{\providecommand{\cmsLeft}{top}}{\providecommand{\cmsLeft}{left}}
\ifthenelse{\boolean{cms@external}}{\providecommand{\cmsRight}{bottom}}{\providecommand{\cmsRight}{right}}
\cmsNoteHeader{FSQ-12-002}
\title{Azimuthal decorrelation of jets widely separated in rapidity in pp collisions at \texorpdfstring{$\sqrt{s} =  7\TeV$}{sqrt(s) = 7 TeV}}

\date{\today}

\abstract{
The decorrelation in the azimuthal angle between the most forward and the most backward jets (Mueller--Navelet jets) is measured in data collected in $\Pp \Pp$ collisions with the CMS detector at the LHC at $\sqrt{s} = 7\TeV$. The measurement is presented in the form of distributions of azimuthal-angle differences, $\Delta\phi$, between the Mueller--Navelet jets, the average cosines of $(\pi-\Delta\phi)$, $2(\pi-\Delta\phi)$, and $3(\pi-\Delta\phi)$, and ratios of these cosines. The jets are required to have transverse momenta, $\pt$, in excess of 35\GeV and rapidities, $\vert y \vert$, of less than 4.7. The results are presented as a function of the rapidity separation, $\Delta{y}$, between the Mueller--Navelet jets, reaching $\Delta{y}$ up to 9.4 for the first time. The results are compared to predictions of various Monte Carlo event generators and to analytical predictions based on the DGLAP and BFKL parton evolution schemes.
}

\hypersetup{%
pdfauthor={CMS Collaboration},%
pdftitle={Azimuthal decorrelation of jets widely separated in rapidity in pp collisions at sqrt(s) =  7 TeV},%
pdfsubject={CMS},%
pdfkeywords={CMS, physics, QCD, jets, dijets, large rapidity separation, BFKL}}

\maketitle
\section{Introduction}

Quantum chromodynamics (QCD), the theory of strong interactions, has been successfully tested in hard processes in high-energy particle collisions. Perturbative QCD calculations performed within the framework of collinear factorisation using the Dokshitzer--Gribov--Lipatov--Altarelli--Parisi (DGLAP) parton evolution scheme~\cite{Gribov:1972ri, Gribov:1972rt, Lipatov:1974qm, Altarelli:1977zs, Dokshitzer:1977sg} have been found to describe many measurements well.

An appropriate tool for QCD studies are hadronic jets---collimated bunches of hadrons, which are the visible manifestations of the energetic partons emerging from the underlying processes. At leading order in the strong coupling \alpS, QCD predicts the production of two partons back-to-back in the azimuthal plane and consequently---even after parton showering and hadronisation---the appearance of two jets with a strong correlation in their azimuthal angle. A deviation from the back-to-back configuration and a weakening of the correlation, namely a decorrelation, occurs if higher-order processes are considered and more partons appear in the final state.

At high centre-of-mass energies, $\sqrt{s} \to \infty$, a kinematical domain can be reached where semi-hard parton interactions with transverse momenta $\pt \ll \sqrt{s}/2$ play a substantial role. This asymptotic domain is more appropriately described by the Balitsky--Fadin--Kuraev--Lipatov (BFKL) evolution equation~\cite{Kuraev:1976ge, Kuraev:1977fs, Balitsky:1978ic} than by the DGLAP approach. In pp collisions, such a regime can be experimentally approached by requiring two low-$\pt$ jets that are widely separated in rapidity, $y$~\cite{Mueller:1986ey}---a scenario for which BFKL, in contrast to DGLAP, predicts a strong rise of the inclusive dijet cross section with increasing rapidity separation. In a kinematic region where semi-hard parton interactions are important, the azimuthal decorrelation will increase with increasing $\Delta{y} = \vert y_1 - y_2 \vert$ between the jets~\cite{DelDuca:1993ga,Stirling:1994he}, where $y_1$ and $y_2$ are rapidities of the most forward and the most backward jets (Mueller-Navelet jets, MN)~\cite{Mueller:1986ey}. The large LHC centre-of-mass energy, and the large pseudorapidity coverage of the detectors, allows multijet production to be explored in a region of $\Delta{y}$ that was previously kinematically inaccessible. The BFKL approach was derived in the infinite-energy limit using the leading-logarithm (LL) approximation. At finite energy, the BFKL approach can be significantly improved using the next-to-leading-logarithm (NLL) approximation~\cite{Fadin:1998py, Ciafaloni:1998gs, Brodsky:1998kn, Ciafaloni:1999yw}, which incorporates further elements like energy-momentum conservation and correlations at small rapidities.

Earlier searches for BFKL signatures in hadron-hadron collisions using events with jets widely separated in rapidity were made at the Tevatron by \DZERO~\cite{Abachi:1996et,Abbott:1999ai}. The \DZERO measurements of azimuthal decorrelation were restricted to a pseudorapidity separation $\Delta\eta<6$, where $\eta=-\log[\tan(\theta/2)]$ and $\theta$ is the polar angle relative to the beam direction. No significant indications of BFKL effects were found~\cite{Abachi:1996et}. Studies~\cite{Abbott:1999ai} have revealed a strong dependence of the dijet production cross section at large rapidity separation on the collision energy. At the LHC, such measurements can be performed at much higher collision energies and with larger rapidity separation between the jets, thus enhancing the possibility to observe BFKL signatures in the data.

Both ATLAS~\cite{Aad:2011jz} and CMS~\cite{Chatrchyan:2012pb} have published measurements of dijet production in pp collisions
at 7\TeV as a function of the rapidity separation between the two jets, and these measurements do not show evidence for BFKL signatures in events with jets with $\pt > 35\GeV$.

Although theoretical arguments support the idea that azimuthal decorrelation observables have greater sensitivity to BFKL effects~\cite{Vera:2007kn}, an earlier ATLAS measurement~\cite{Aad:2014pua} between leading and subleading jets with $\pt > 60\GeV$ and $\pt > 50\GeV$ and $\Delta{y} < 8$ did not indicate any deviation from DGLAP predictions. However, the analysis of MN dijets with $\pt > 35\GeV$ and $\Delta{y} < 9.4$ presented in this paper should be more sensitive to BFKL effects, given the lower jet $\pt$ thresholds and the wider dijet rapidity separation considered.
Studies of jets with large rapidity separation require data collected at low instantaneous luminosity to avoid contamination from jets produced in different overlapping pp collisions~\cite{Chatrchyan:2012pb}. In this paper, observables connected to the azimuthal decorrelation of MN dijets are presented that use a data sample corresponding to an integrated luminosity of ${\approx}41\pbinv$ collected during proton-proton running at $\sqrt{s}=7\TeV$ in the year~2010.

\section{Physics motivation and Monte Carlo event generators}

The normalised cross section as a function of the azimuthal-angle difference ($\Delta\phi$) between MN jets  with $\pt  > p_{\mathrm{T min}}$ can be written as a Fourier series~\cite{DelDuca:1993ga, Stirling:1994he}:
\begin{equation}
\frac{1}{\sigma} \frac{\rd \sigma}{\rd (\Delta \phi) } (\Delta{y}, p_{\mathrm{T min}})   =
\frac{1}{2 \pi} \biggl[ 1 + 2 \sum_{n=1}^{\infty}  C_n (\Delta{y}, p_{\mathrm{T min}}) \, \cos(n (\pi - \Delta \phi)) \biggr].
\end{equation}
The Fourier coefficients $C_n$  are equal to the average cosines of the decorrelation angle, $( \pi - \Delta \phi)$:
$C_n (\Delta{y}, p_{\mathrm{T min}}) = \langle\cos(n ( \pi - \Delta \phi))\rangle$, where $ \Delta \phi = \phi_1 - \phi_2 $
is the difference between the azimuthal angles $\phi_1$ and $\phi_2$ of the MN jets.

If there are only two jets in the final state, they have to be approximately back-to-back in the azimuthal plane ($ \Delta \phi = \pi $)
and the average cosines equal unity: $\langle\cos(n(\pi - \Delta \phi))\rangle = 1$. Due to parton radiation, the $(\pi - \Delta \phi)$ distribution has a non-zero width that is determined by Fourier harmonics involving $\langle\cos(n(\pi - \Delta \phi))\rangle $. In the BFKL approach, an increasing rapidity interval between the MN jets leads to an increased number of emitted partons and thus to an increased azimuthal decorrelation: $\langle\cos(n(\pi - \Delta \phi))\rangle<1$. In the DGLAP picture within the LL approximation, in contrast, the partons emitted between the MN jets have much lower $\pt$ than the latter, and their emission does not depend on their rapidity separation.
Hence, parton emissions from the parton cascade can change the azimuth of the parent partons to a much lesser extent than in the BFKL approach where the $\pt$ of mother and daughter partons can be very similar. However, when the MN jets are not the jets with the highest $\pt$, then even in the DGLAP picture a significant decorrelation might be observed.

In this paper the average cosines of the azimuthal angle between MN jets,
 $(\pi-\Delta\phi)$, $2(\pi-\Delta\phi)$, and $3(\pi-\Delta\phi)$ (i.e. $C_1$, $C_2$, and $C_3$) are measured as functions of the rapidity separation, $\Delta{y}$, as suggested in Refs.~\cite{DelDuca:1993ga, Stirling:1994he, Kim:1995zu,Vera:2007kn,Angioni:2011wj,Colferai:2010wu,Ducloue:2012bm}. In addition, the ratios of the average cosines $C_2/C_1$ and $C_3/C_2$ are measured, as proposed in Refs.~\cite{Vera:2007kn,Angioni:2011wj,Colferai:2010wu,Ducloue:2012bm}. To cover all available $\Delta{y}$ space, $\Delta\phi$ distributions are measured in three bins of rapidity separation: $\Delta{y}<3.0$, $3.0<\Delta{y}<6.0$, and $6.0<\Delta{y}<9.4$.
The average cosines may be expressed explicitly using conformal symmetries of the BFKL evolution equation~\cite{Brodsky:1998kn},
which are absent in the DGLAP evolution equation. Moreover, since one expects a suppression of DGLAP contributions in the two ratios~\cite{Angioni:2011wj}, they are particularly sensitive to manifestations of BFKL effects. In addition, uncertainties related to the factorisation and renormalisation  scales are reduced in the ratios~\cite{Ducloue:2013hia}.

The measurements are performed with the CMS detector, using proton-proton collision data recorded at $\sqrt{s}=7\TeV$ for jets with $\pt>35\GeV$ and $\vert y \vert<4.7$, allowing a rapidity separation between the MN jets of up to $\Delta{y}=9.4$. The jets are reconstructed with the anti-\kt algorithm~\cite{AntiKT, AntiKTimpl} with a distance parameter $R=0.5$.

The measured jet observables, corrected to the stable-particle level (lifetime $c\tau > 1\unit{cm}$), are compared to predictions from various Monte Carlo (MC) event generators which extend the DGLAP approach by including LL soft and collinear radiation in their parton-shower modelling: \PYTHIA~6 (version 6.422)~\cite{Pyt} tune Z2~\cite{PytZ2}, \textsc{herwig}++ (version 2.5.1) tune UE-7000-EE-3~\cite{Hw++}, and \PYTHIA~8 (version~8.145)~\cite{Pyt8} tune 4C~\cite{Corke:2010yf}. In the mentioned generators, different models are used for the simulation of multiparton interactions and hadronisation. The parameters of multiparton interactions in these tunes are adjusted to best describe LHC data. The MC generator \POWHEG~\cite{Nason:2004rx,Frixione:2007vw,Alioli:2010xd}---using the \textsc{cteq6m} parton distribution function~\cite{Pumplin:2002vw}, and interfaced with \PYTHIA~6 and 8---is used to investigate the sensitivity of the measured jet observables to the contribution of next-to-leading-order (NLO) terms. The measurements are also compared to the DGLAP-based MC generator \SHERPA 1.4~\cite{Gleisberg:2008ta}, which uses tree-level matrix elements for $2 \rightarrow 2+n$-jets (with $n = 0$, $1$, $2$ in this work) matched to LL parton showers. Finally, data-theory comparisons are also performed using the analytical NLL BFKL predictions as obtained in Ref.~\cite{Ducloue:2013bva} at parton level, as well as with predictions obtained from the \textsc{hej+ariadne} generator package (version 0.99b)~\cite{hejAriadne}. The latter consists of \textsc{hej} version 1.3.2~\cite{HEJ}, which is based on LL BFKL matrix elements, and the hadronisation and parton-shower package of \textsc{ariadne} 4.12~\cite{ARIADNE}.

\section{The CMS detector}

The most relevant component of the CMS detector~\cite{cmsexp} for this analysis is the calorimeter system, which covers the pseudorapidity range $\abs{\eta} < 5.2$. The crystal electromagnetic calorimeter (ECAL) and the brass/scintillator hadron calorimeter (HCAL) extend to $\abs{\eta} = 3.0$. The HCAL cells map to an array of ECAL crystals to form calorimeter towers projecting radially outwards from the nominal interaction point. The pseudorapidity region $3.0<\abs{\eta}<5.2$ is covered by the hadronic forward (HF) calorimeter, which consists of steel absorber wedges with embedded radiation-hard quartz fibres, oriented parallel to the beam direction. The calorimeter towers in the barrel region have a segmentation of $\Delta\eta{\times}\Delta \phi = 0.087{\times}0.087$, becoming progressively larger in the endcap and forward regions ($\Delta\eta{\times}\Delta \phi = 0.175{\times}0.175$ at $\eta \approx 4.5$).

The silicon tracker measures charged particles within the pseudorapidity range $\abs{\eta}< 2.5$. It consists of 1440 silicon pixel and 15\,148 silicon strip detector modules and is, like ECAL and HCAL, located in the 3.8\unit{T} field of the superconducting solenoid. It provides an impact parameter resolution of 50--175\mum~\cite{TRK-11-001pub}, and thus, precise interaction vertex reconstruction using charged particle tracks within its acceptance.

The CMS trigger system consists of a hardware level-1 trigger and a software high-level trigger. Jets formed online by the trigger system use ECAL, HCAL, and HF inputs for energy clustering and are not corrected for the jet energy response.

\section{Event selection}

Dijet events with a large rapidity separation are rare.
Therefore, in addition to the standard single-jet trigger that selects events containing at least one jet with raw $\pt > 15\GeV$, a dedicated trigger for forward-backward dijets was implemented that selects events with two jets in opposite hemispheres, each with $\abs{\eta}>3.0$ and jet raw $\pt > 15\GeV$.
In order to keep the rate of the single-jet trigger within the allocated bandwidth, a prescale factor of ${\approx}10^{3}$ was used, and an effective integrated luminosity of ${\approx}33\nbinv$ is recorded with it.
The forward-backward trigger was operated with a moderate prescale factor of ${\approx}8$, recording an effective integrated luminosity of ${\approx}5\pbinv$, resulting in the collection of a sample of large $\Delta{y}$ dijet events ($\Delta{y} > 6$), 100 times larger than that collected with the single-jet trigger alone.

The single-jet trigger efficiency is measured by means of a control sample selected
with the minimum-bias trigger, which maximises the data collection efficiency while maintaining a low background level~\cite{QCD-07-002}. The single-jet trigger is measured to be 99.5\% efficient for events containing dijets with  $\pt > 35\GeV$ and is used for the determination of the efficiency of the forward-backward dijet trigger. The latter is measured to be 100\% efficient for dijets with $\pt > 35\GeV$.

Jets are reconstructed offline using the energy depositions in the calorimeter towers.
In the reconstruction process, the contribution from each tower is assigned a momentum. The magnitude and the direction of the momentum are given by the energy measured in the tower and the coordinates of the tower, respectively. The raw jet energy is obtained from the sum of the tower energies, while the raw jet momentum is calculated from the vectorial sum of the tower momenta. The raw jet energies are then corrected to establish a uniform relative response of the calorimeter in $\eta$ and a calibrated absolute response in $\pt$~\cite{JME-10-011pub}. The jet energy resolution (JER) for calorimeter jets with $\pt \approx 35\GeV$ is about $22\%$ for $\vert \eta \vert <0.5$ and about $10\%$ for $4< \vert \eta \vert <4.5$~\cite{JME-10-003}. The uncertainty on the jet energy calibration for jets with $\pt \approx 35\GeV$ depends on $\eta$ and is
${\approx}$7--8$\%$~\cite{JME-10-011pub}.

In order to reduce the sensitivity to overlapping pp collisions within a single bunch crossing (so-called ``pileup'' events), only events with exactly one reconstructed pp primary vertex within the luminous region are used for the measurement. This selection leads to about 30\% events lost, whereas without this selection the average number of pileup interactions over analysed data was ${\approx}2.2$~\cite{Chatrchyan:2012gwa}. The primary vertex is required to be reconstructed within ${\pm}24\unit{cm}$ of the nominal interaction point along the beamline~\cite{TRK-11-002pub}.

Loose jet quality requirements~\cite{JME-09-008} are applied to suppress the effect of calorimeter noise. Events with at least two jets with $\pt > 35\GeV$ and $\abs{y}< 4.7$ are selected, and only jets satisfying these criteria are used for the analysis.

Mueller--Navelet jet pairs are constructed from jets passing the above criteria.
The azimuthal-angle difference $\Delta\phi$ between the two jets is measured in the range $0<\Delta\phi<\pi$ for three bins of rapidity separation between the MN jets: $ \Delta{y}  < 3.0$, $3.0 < \Delta{y} < 6.0$, and $6.0 <  \Delta{y} < 9.4$, normalised to unity integral. The average cosines $C_1$, $C_2$, and $C_3$
are measured in bins of $\Delta{y}$ up to 9.4. The cosine ratios $C_3/C_2$ and $C_2/C_1$ are calculated as ratios of average cosines for each bin in $\Delta{y}$.

\section{Corrections for detector effects}

The finite jet $\pt$ resolution results in jet $\pt$ values at the detector level that deviate from those at stable-particle level. Due to the steep slope of the $\pt$ spectrum, jets with smaller $\pt$ may migrate to higher $\pt$ and thus increase the number of jets in distributions at the detector level. The finite jet $\eta$ resolution and measurement offset lead to a finite $\Delta{y}$ resolution and offset, such that dijets may migrate from one $\Delta{y}$ bin to another. Similarly, distributions in $\Delta \phi$ are affected by the finite $\phi$ resolution.

These effects are mitigated using corrections derived with a hybrid method. This method comprises both a multiplicative correction designed to compensate migrations in the jet $\pt$ space and a full unfolding in the ($\Delta{y}$, $\Delta \phi$) space.
The migration of jets into and out of the analysed phase space leads to the selection of different dijets at detector and stable-particle level, which is accounted for as a non-negligible background and a limited detection efficiency. Bin-wise multiplicative corrections for background and efficiency are derived from MC simulation and applied to the distributions before and after correction for inter-bin migrations, respectively. The associated purity is always larger than 91\%. To account for inter-bin migrations, the $\Delta\phi$ distributions are unfolded with an iterative procedure~\cite{DAgostini:1994zf} in each of the three analysed $\Delta{y}$ bins.
Probabilities for inter-bin migration were calculated for all $\Delta\phi$ bins in all three $\Delta{y}$ ranges, and they were found to always be less than 20\%. The same unfolding procedure is applied to the 2-dimensional ($\Delta{y}$, $\Delta \phi$) distributions for the calculation of $\langle\cos(n(\pi - \Delta \phi))\rangle$ at the stable-particle level. The correction factors associated with the hybrid method were found to be 0.6--1.1 for the $\Delta\phi$ distributions and 0.9--1.05 for $\langle\cos(n(\pi - \Delta \phi))\rangle$.

The corrections are calculated from the simulated events generated with \PYTHIA 6 (version 6.422, Z2 tune) and \HERWIGpp (version 2.4.1, default tune). These events are passed through the full CMS detector simulation based on \GEANTfour~\cite{geant4}. The averages of the corrected values obtained using \PYTHIA 6 and \HERWIGpp are taken as the final, corrected values of the observables.
The hybrid unfolding procedure is tested by comparing the MC prediction at the stable-particle level with the distribution corrected from 
detector level to stable-particle level. Closure tests, where distributions generated with \PYTHIA 6 (\HERWIGpp) are unfolded with a response 
matrix obtained from \PYTHIA 6 (\HERWIGpp), show differences of less than 1\%. Cross-closure tests, i.e. unfolding \PYTHIA 6 distributions with 
a response matrix from \HERWIGpp and vice versa, show differences of less than 7\%, which is much smaller than the differences between 
the stable-particle distributions predicted by both models.

In Ref.~\cite{JME-10-011pub} it was shown that the jet energy resolution (JER) for calorimeter jets in the simulation is 6.5--14.9\% better than the one found in data. To correct for this discrepancy, an additional smearing was applied to detector-level jets in the MC simulation.

\section{Experimental uncertainties}

The systematic uncertainties of the measurement are evaluated in the following way:

-- To calculate the effect of the jet energy scale (JES) uncertainty, the $\pt$ values of the jets are varied by $\pt$-dependent and $\eta$-dependent values~\cite{JME-10-011pub}. Observables were then recalculated twice---with the $\pt$ values varied up and down---and the difference between the results defines the uncertainty of the observable associated with JES.

-- The JER obtained in MC simulations differs from that observed in data~\cite{JME-10-011pub} (as discussed at the end of Section 5), while the uncertainty of the discrepancy varies between $7.6\%$ and $23.7\%$, depending on $\eta$. The impact of this uncertainty is assessed by varying, in the MC simulation, the amount of $\pt$ smearing on detector-level jets. The difference between the results again defines the uncertainty.

-- The sensitivity of the measurement to pileup is investigated using collision data. In the analysis the number of primary vertices per event is required to be equal to 1. However, as the primary vertex reconstruction is not $100\%$ efficient, a residual dependence of observables on pileup may be present. The available data are divided into two sets corresponding to different instantaneous bunch luminosities. In one set, the average number of pp interactions was restricted to be less than two, while in the other set more than two pp interactions in average were required. The observables obtained from each set are compared, and no dependence on the instantaneous bunch luminosity is found.

-- The uncertainty of data correction to the stable-particle level (see Section 5) is determined from \PYTHIA 6 and \HERWIGpp. The difference between the corrections obtained with the two different MC generators is taken as the systematic uncertainty for the model dependence, and it never exceeds $6.4\%$ together with the uncertainty due to limited MC statistics being added in quadrature.

-- In order to estimate the impact of the imprecise modelling of the angular resolution for jets in the MC simulation, an extra smearing is applied to the difference between the jets' azimuthal separation at the detector level and at the stable-particle level. This difference is varied by ${\pm}10\%$~\cite{JME-10-003}, which is a conservative estimation of the real smearing, and the same procedure is performed for the $\eta$ difference. The resulting change in the measurements turns out to be negligible and is not included in the systematic uncertainty.

The total systematic uncertainty of the measurement is obtained by quadratically summing the individual experimental uncertainties listed above. The individual contributions to the total uncertainty are summarised in Table 1, together with the statistical uncertainties. The ranges correspond to the variation of the uncertainty with $\Delta \phi$ or with $\Delta{y}$, and for asymmetric uncertainties the upper and lower limits are shown.

\begin{table}[htbp]
\begin{center}
\topcaption{Systematic and statistical uncertainties (\%) of the observables measured in this work.
}\label{tabunc}
\begin{tabular}{c|ccccc}
\hline
Observable & JES & JER & Corrections  & Total systematic & Statistical\\
\hline
$\Delta\phi (\vert \Delta{y} \vert <3.0)$ &  $^{+(2.3 - 13.7)}_{-(3.0 - 10.2)}$ &  $^{+(0.1 - 10.6)}_{-(0.4 - 7.6)}$ & 0.1--2.0 & $^{+(2.3 - 17.4)}_{-(3.0 - 12.7)}$ & 0.3--5.1\phantom{0} \\[6pt]
$\Delta\phi (3.0 < \vert \Delta{y} \vert < 6.0)$ & $^{+(2.5 - 16.4)}_{-(2.9 - 10.8)}$ & $^{+(0.7 - 6.2)}_{-(0.8 - 3.4)}$ & 0.4--2.3 & $^{+(3.0 - 17.5)}_{-(3.1 - 11.3)}$ & 0.9--6.2\phantom{0} \\[6pt]
$\Delta\phi (6.0 < \vert \Delta{y} \vert < 9.4)$ &  $^{+(2.1 - 31.5)}_{-(1.9 - 17.3)}$ & $^{+(5.8 - 17.4)}_{-(2.1 - 9.7)}$  & 0.4--4.5 &  $^{+(6.8 - 32.6)}_{-(3.6 - 19.5)}$ & 5.3--22.0 \\[6pt]
$C_1$  & 1.0--5.5\phantom{0} & 0.6--4.6\phantom{0} & 0.1--3.2 & 1.1--6.5\phantom{0} & 0.2--9.7\phantom{0}\\[1pt]
$C_2$  & 1.8--16.9 & 1.0--4.0\phantom{0} & 0.1--4.9 &  2.3--17.4 & 0.5--17.7\\[1pt]
$C_3$  & 2.7--23.8 & 1.5--15.1 & 0.1--6.4 &  3.2--24.6 & 0.7--23.7\\[1pt]
$C_2/C_1$  & 0.8--12.5  & 0.4--5.6\phantom{0} & 0.1--2.6 & 1.0--13.1 & 0.5--19.7\\[1pt]
$C_3/C_2$  & 0.7--7.1\phantom{0}  & 0.2--7.0\phantom{0} & 0.03--4.3\phantom{0} & 0.7--10.6 & 0.8--28.1\\
\hline
\end{tabular}
\end{center}
\end{table}

\section{Results}

\begin{figure}[hbtp]
  \begin{center}
    \includegraphics[width=0.49\textwidth]{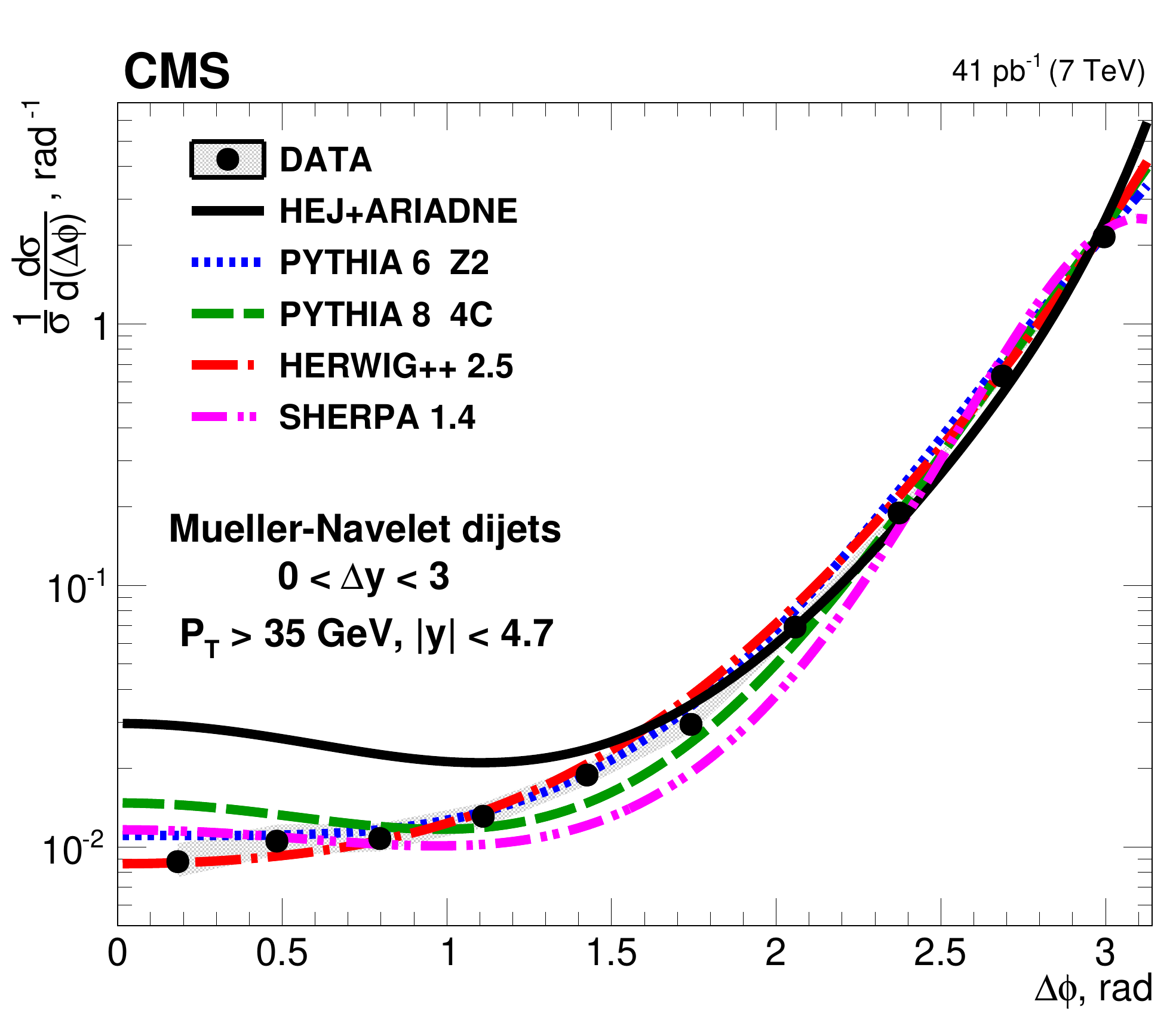}
    \includegraphics[width=0.49\textwidth]{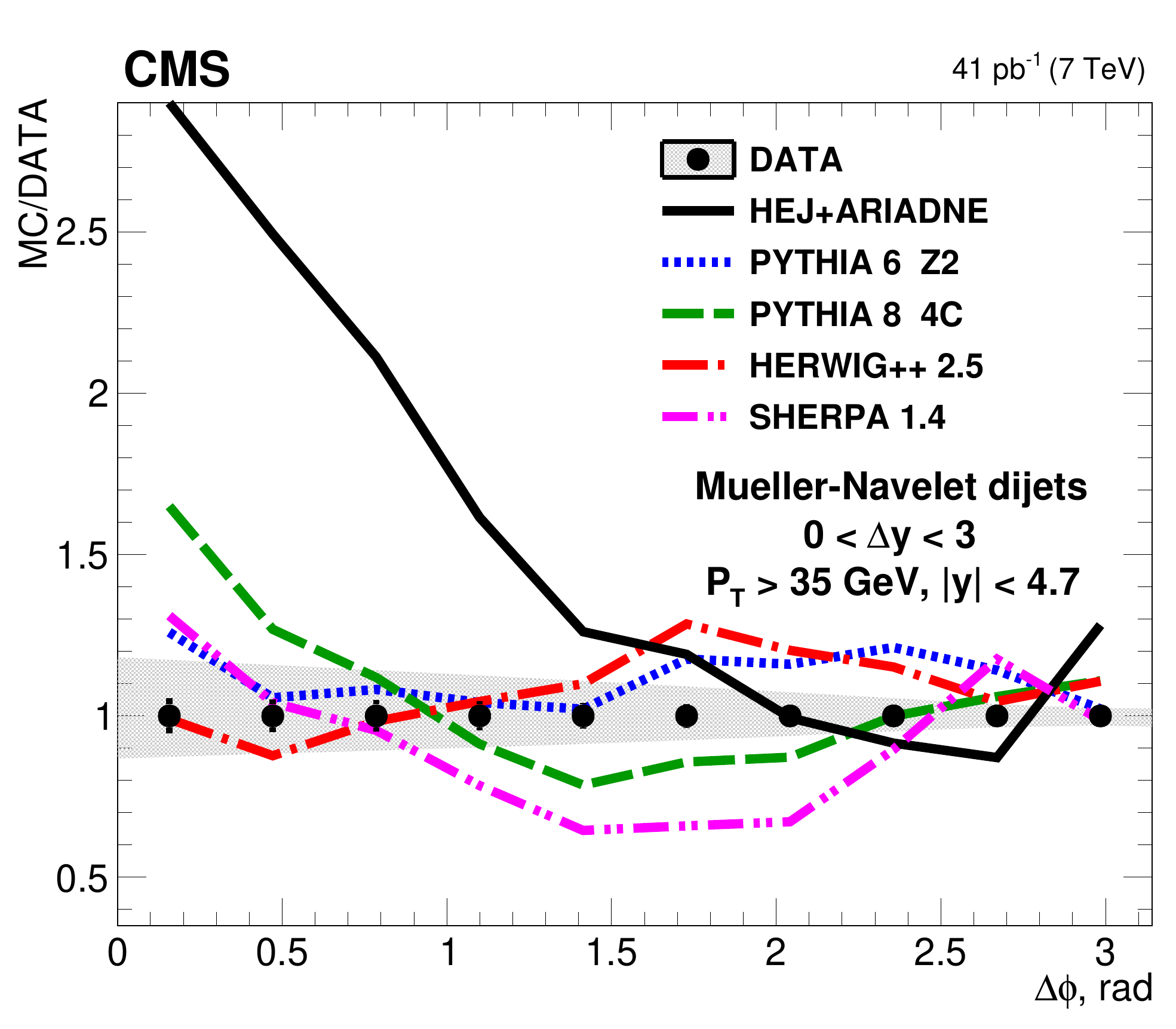}
    \includegraphics[width=0.49\textwidth]{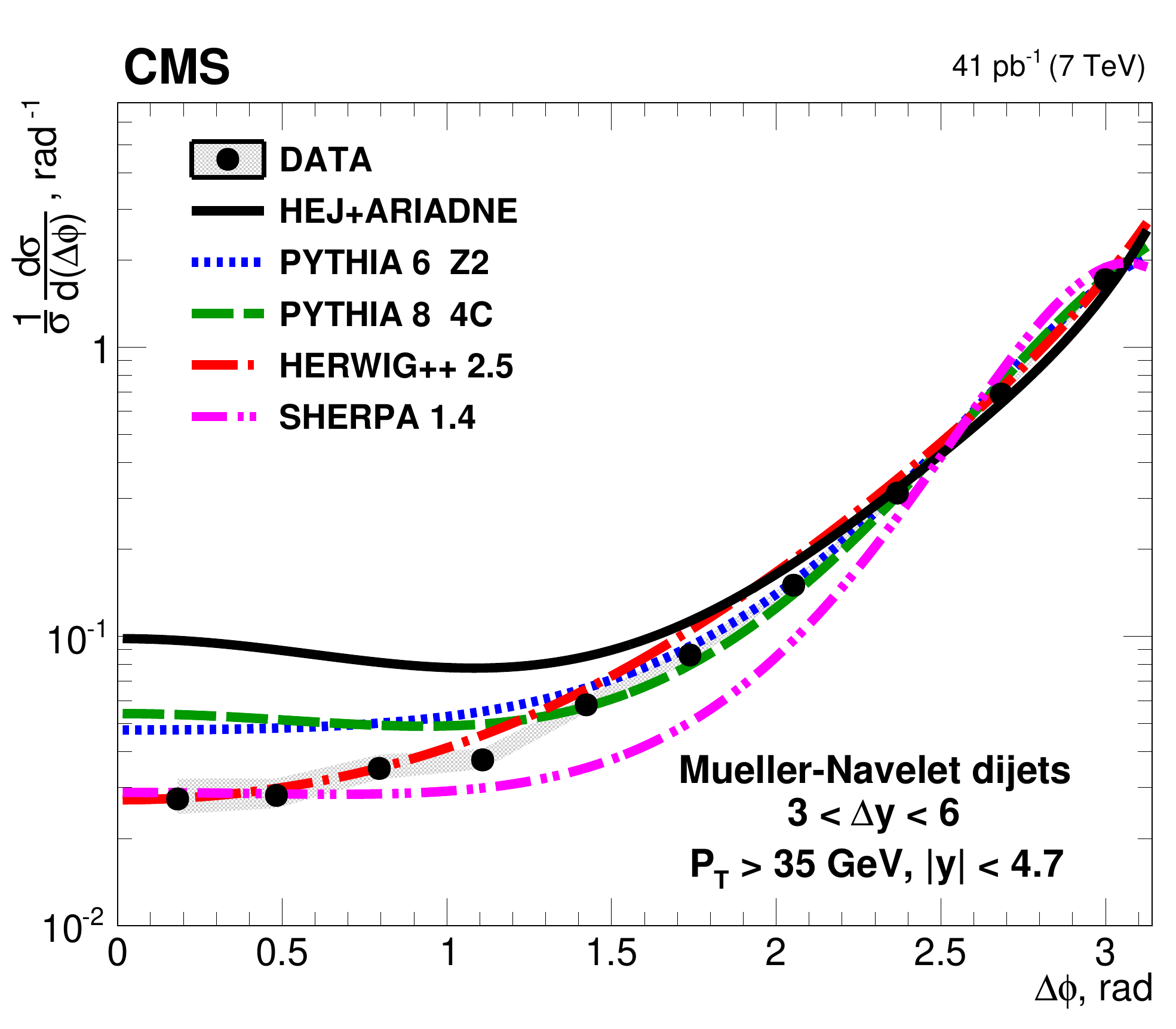}
    \includegraphics[width=0.49\textwidth]{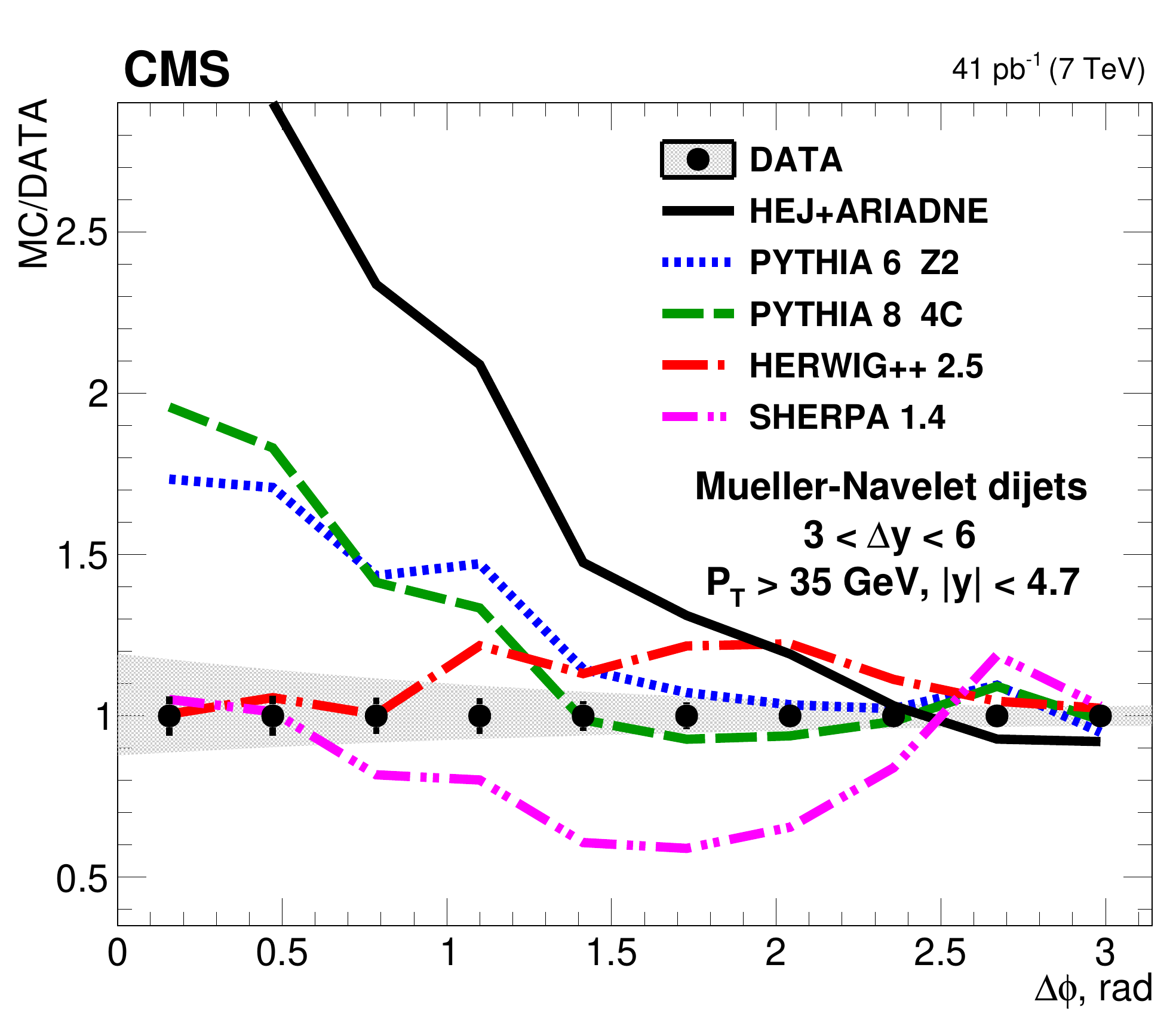}
    \includegraphics[width=0.49\textwidth]{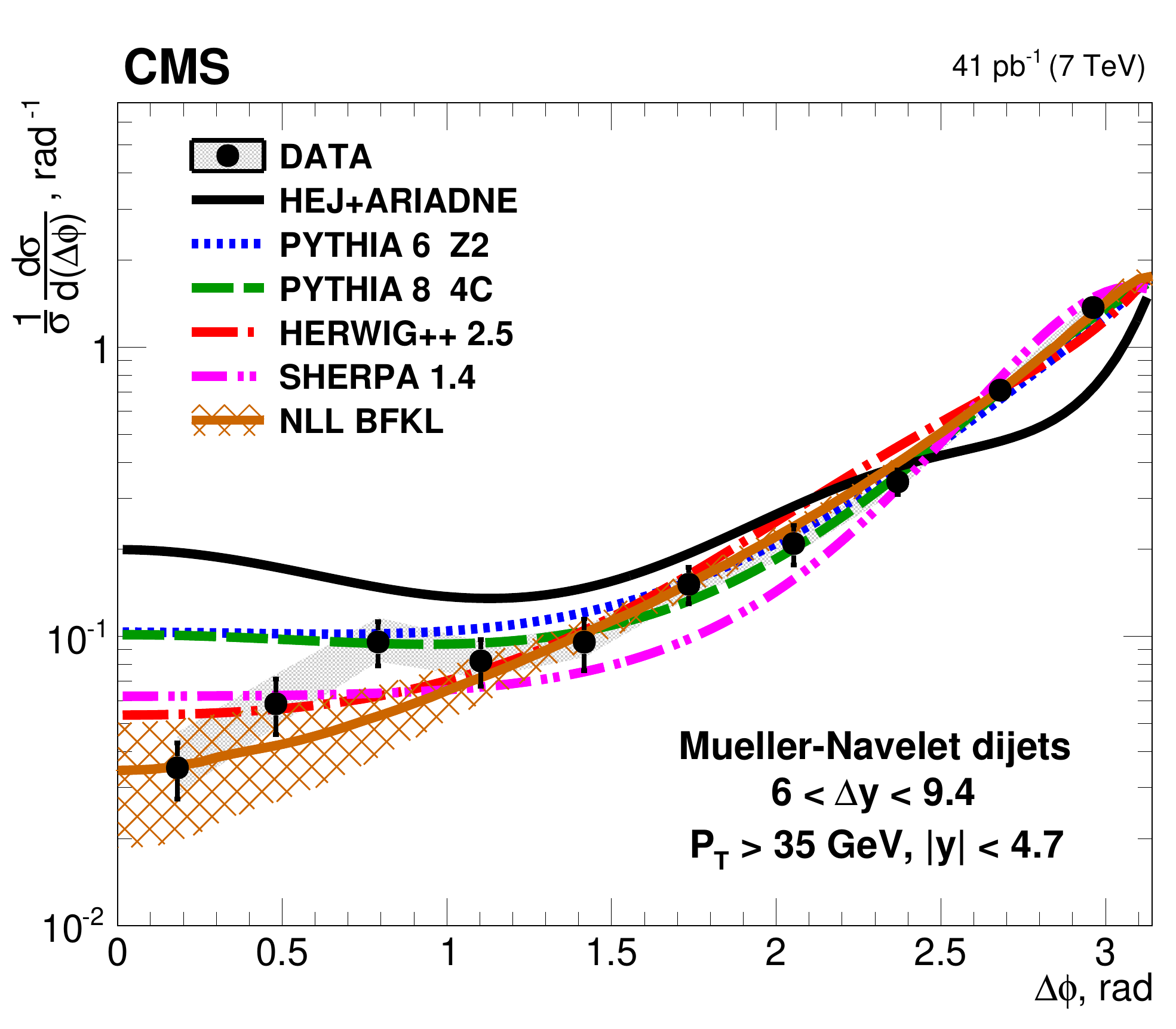}
    \includegraphics[width=0.49\textwidth]{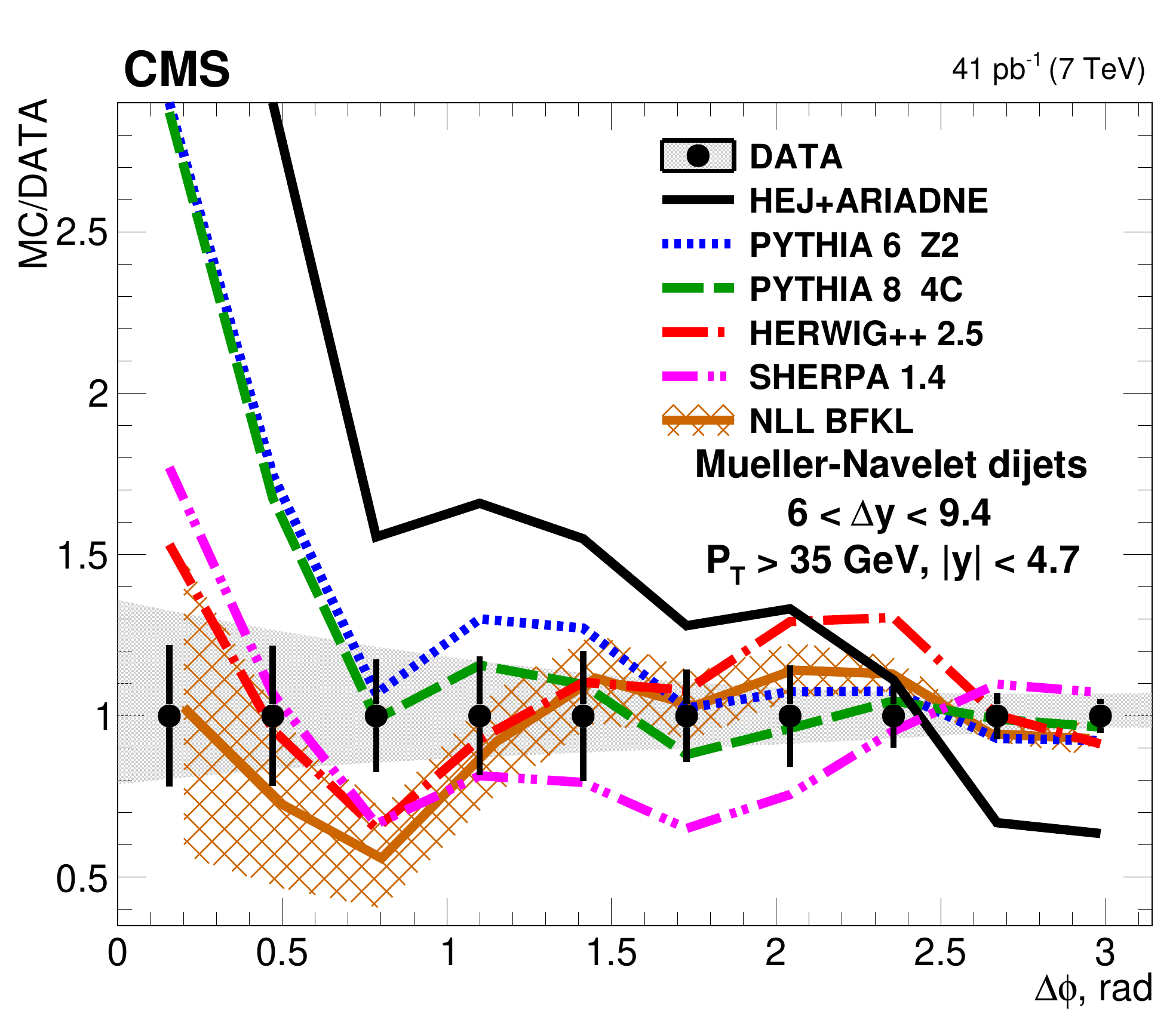}
    \caption{Left: Distributions of the azimuthal-angle difference, $\Delta \phi$, between MN jets in the rapidity intervals $ \Delta{y}  < 3.0$ (top row), $3.0 < \Delta{y}  < 6.0$ (centre row), and $6.0 < \Delta{y}  < 9.4$ (bottom row). Right: Ratios of predictions to the data in the corresponding rapidity intervals. The data (points) are plotted with experimental statistical (systematic) uncertainties indicated by the error bars (the shaded band), and compared to predictions from the LL DGLAP-based MC generators \PYTHIA 6, \PYTHIA 8, \HERWIGpp, and \SHERPA, and to the LL BFKL-motivated MC generator \textsc{hej} with hadronisation performed with \textsc{ariadne} (solid line).}
    \label{fig:MN_dphi_0-3}
    \label{fig:MN_dphi_3-6}
    \label{fig:MN_dphi_6-94}
  \end{center}
\end{figure}

The $\Delta\phi$ distributions for MN dijets measured in the three rapidity intervals $\Delta{y}<3.0$, $3.0<\Delta{y}<6.0$, and  $6.0<\Delta{y}<9.4$ are shown in the left panes of Fig.~\ref{fig:MN_dphi_0-3}. On the right-hand side of Fig.~\ref{fig:MN_dphi_0-3}, the predictions are shown normalised to the data.

The systematic uncertainties are shown as a band around the data points. The measurement shows a high level of back-to-back correlation in the $\Delta{y}<3.0$ bin (Fig.~\ref{fig:MN_dphi_0-3}, top row), while the $\Delta\phi$ distributions become less peaked at $\Delta\phi\approx\pi$ when going to larger $\Delta{y}$ separation (Fig.~\ref{fig:MN_dphi_3-6}, centre and bottom rows). This demonstrates that higher-order corrections at larger $\Delta{y}$ manifest themselves through additional hard-parton radiation.

In the central rapidity interval $\Delta{y}<3.0$ (Fig.~\ref{fig:MN_dphi_0-3}, top row), the LL DGLAP-based MC generators \PYTHIA 6 and \HERWIGpp describe the data well, showing some deviation only at low $\Delta\phi$ values. The LL DGLAP-based MC generators \PYTHIA 8 and \SHERPA, with parton matrix elements matched to LL DGLAP parton showers, exhibit significant deviations from the data beyond the experimental uncertainties at intermediate and large $\Delta\phi$. At intermediate  ($3.0 < \Delta{y}< 6.0$)  and  large ($6.0 < \Delta{y} < 9.4$) rapidity separation, \PYTHIA 6 and 8 show a significant deviation at small $\Delta\phi$ while the measurements are reasonably well described in the region $\Delta\phi > 1.5$. On the contrary, \HERWIGpp and \SHERPA show deviations to the measurements in the medium $\Delta\phi$ region, but are close to the data at very small $\Delta\phi$. The \textsc{hej+ariadne} package overestimates the azimuthal decorrelation at small $\Delta\phi$ at all $\Delta{y}$, though there are a lack of MC data for $6.0 < \Delta{y} < 9.4$. In Fig.~\ref{fig:MN_dphi_0-3} (bottom row) the $\Delta\phi$ distributions are also compared to analytical NLL BFKL calculations at the parton level~\cite{Ducloue:2013bva}, and this comparison is summarised at the end of Section~7, together with the discussion of the other measured observables.

\begin{figure}[hbtp]
  \begin{center}
     \includegraphics[width=0.49\textwidth]{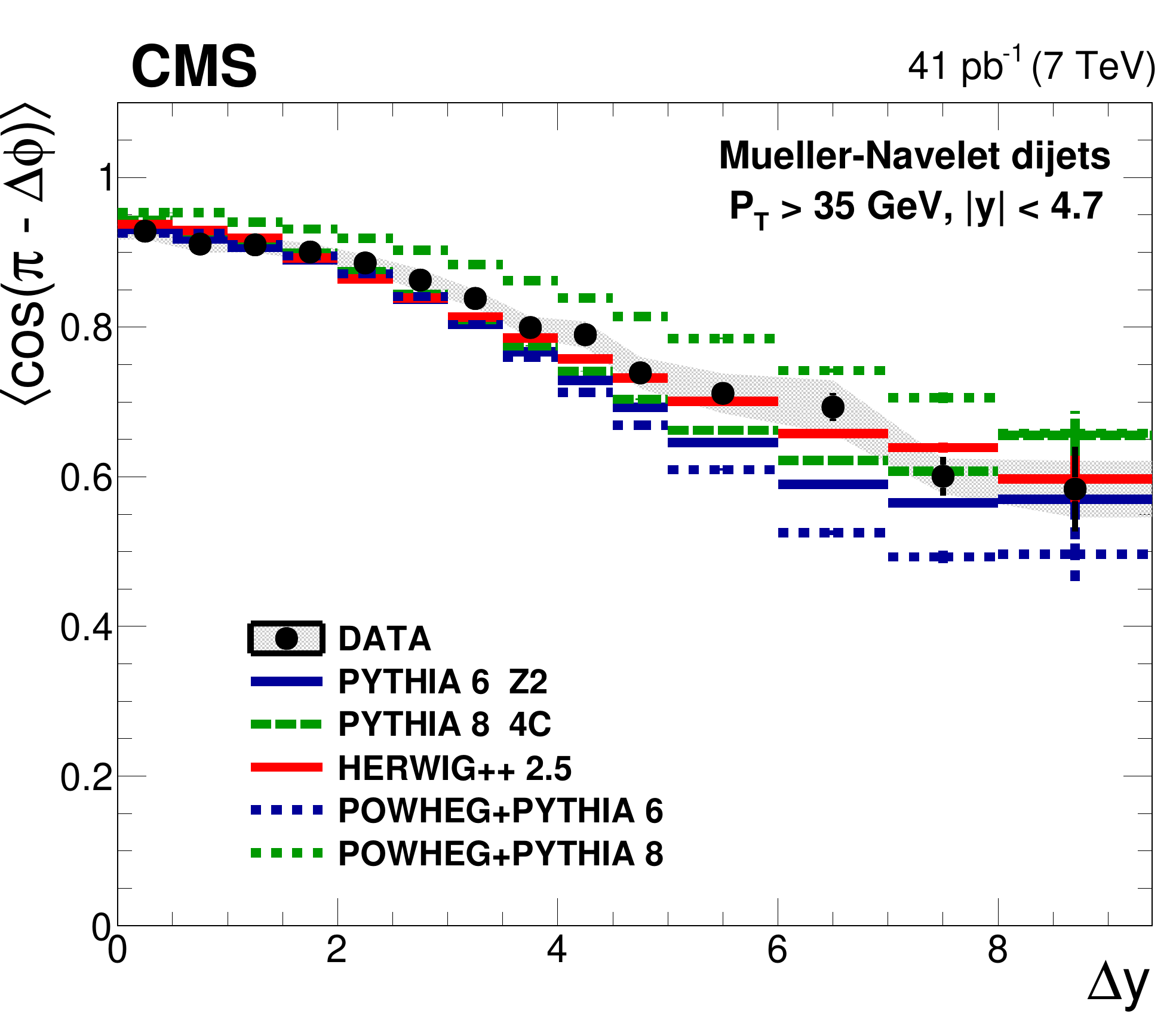}
     \includegraphics[width=0.49\textwidth]{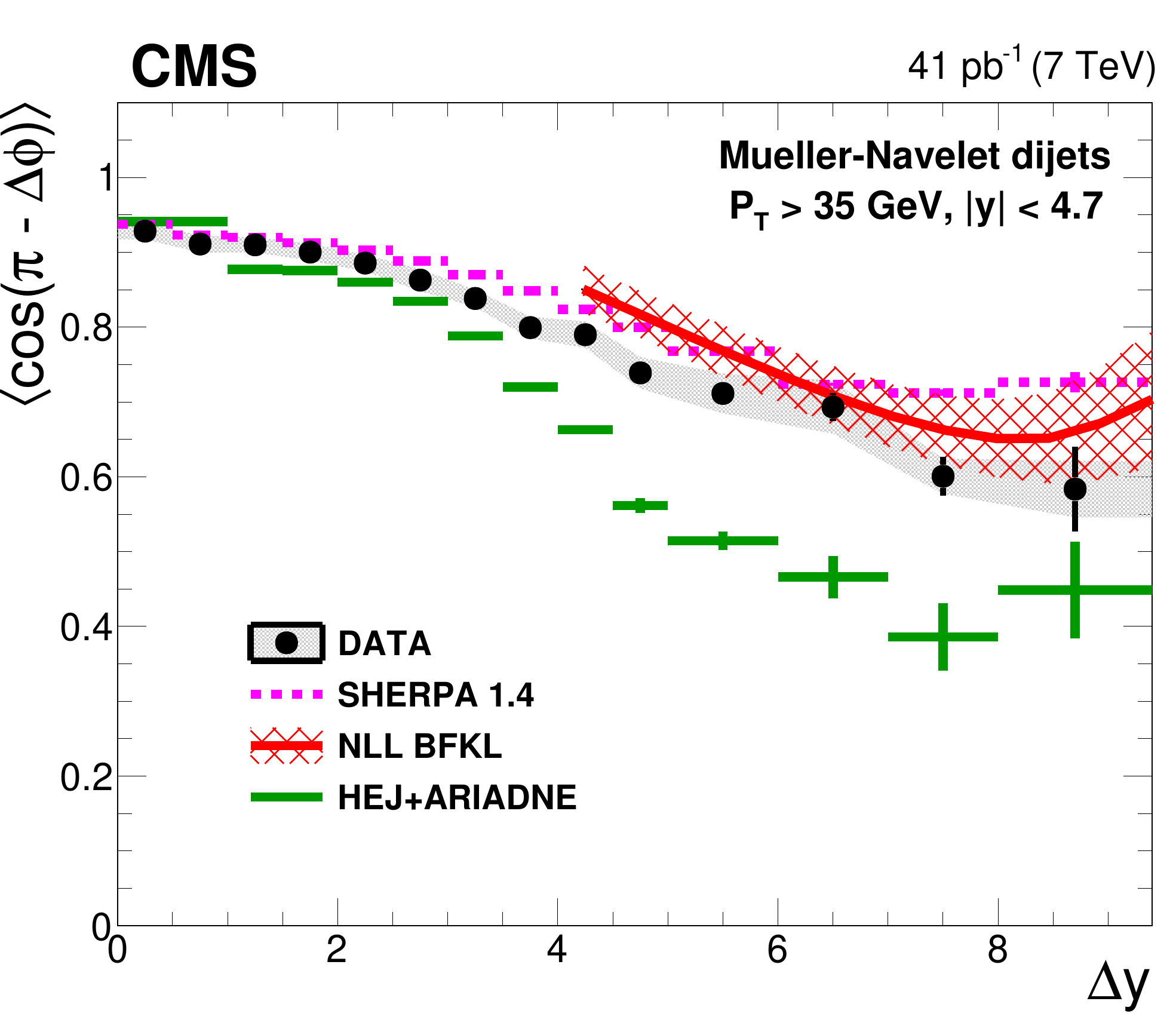}
     \includegraphics[width=0.49\textwidth]{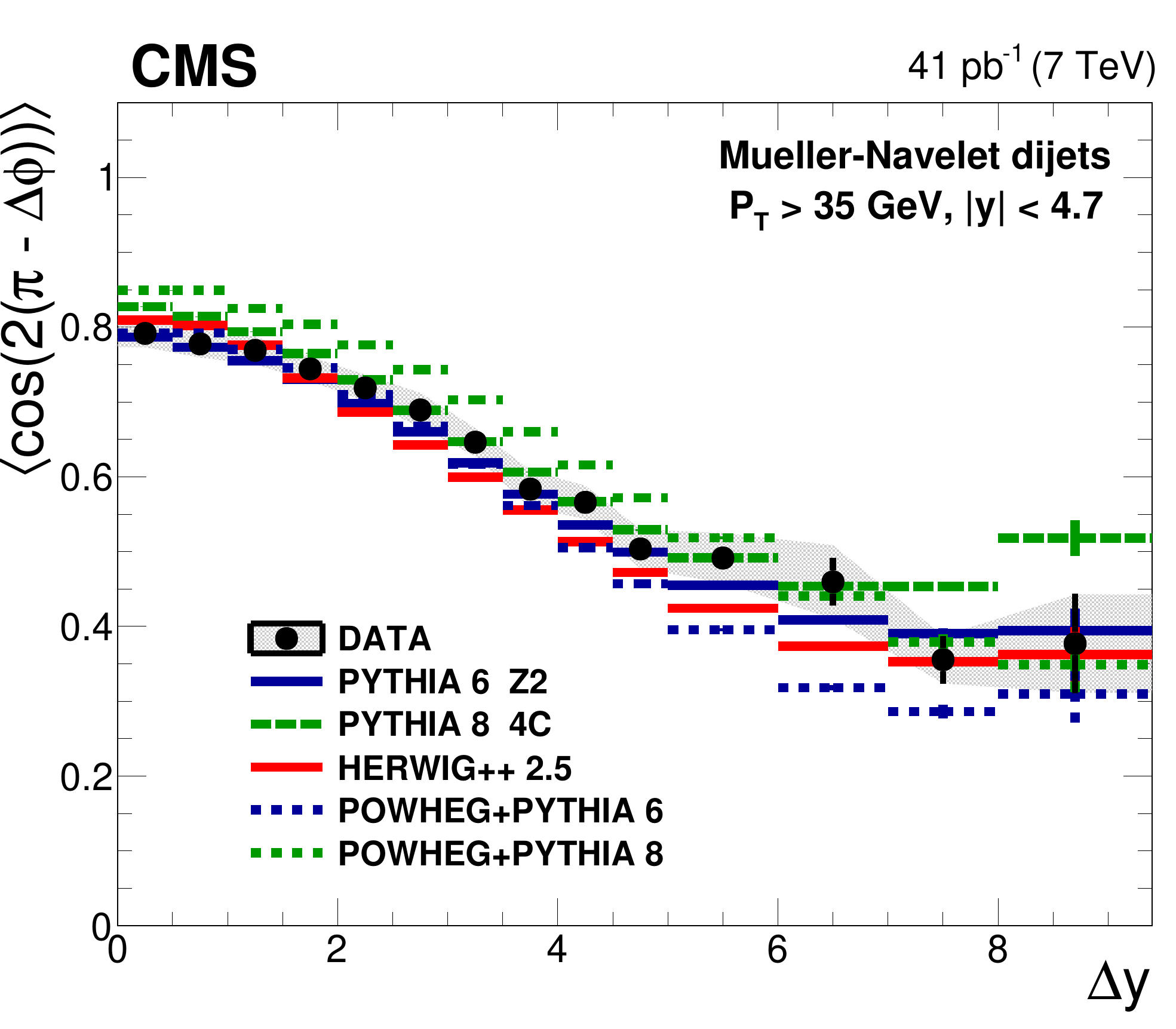}
     \includegraphics[width=0.49\textwidth]{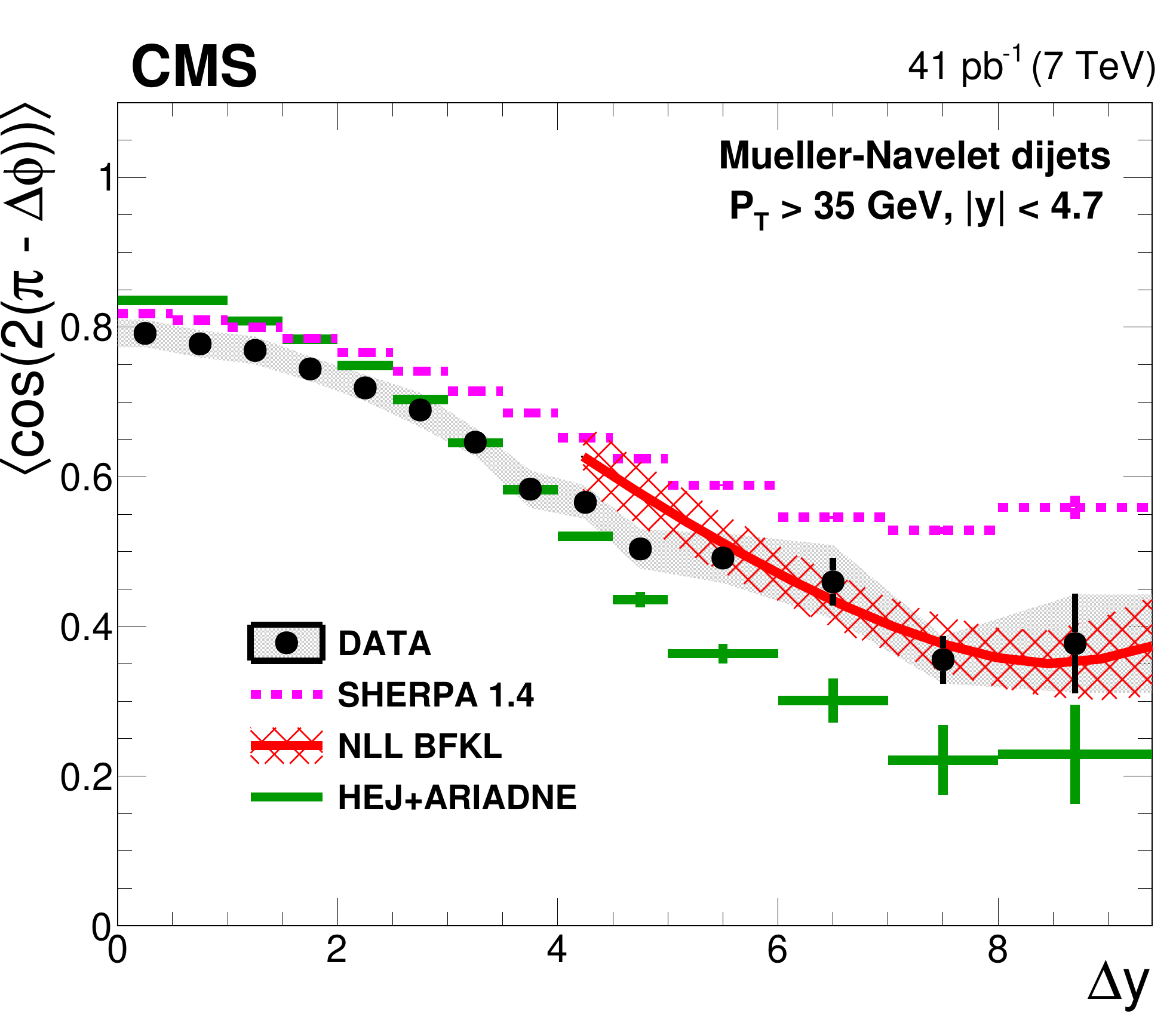}
     \includegraphics[width=0.49\textwidth]{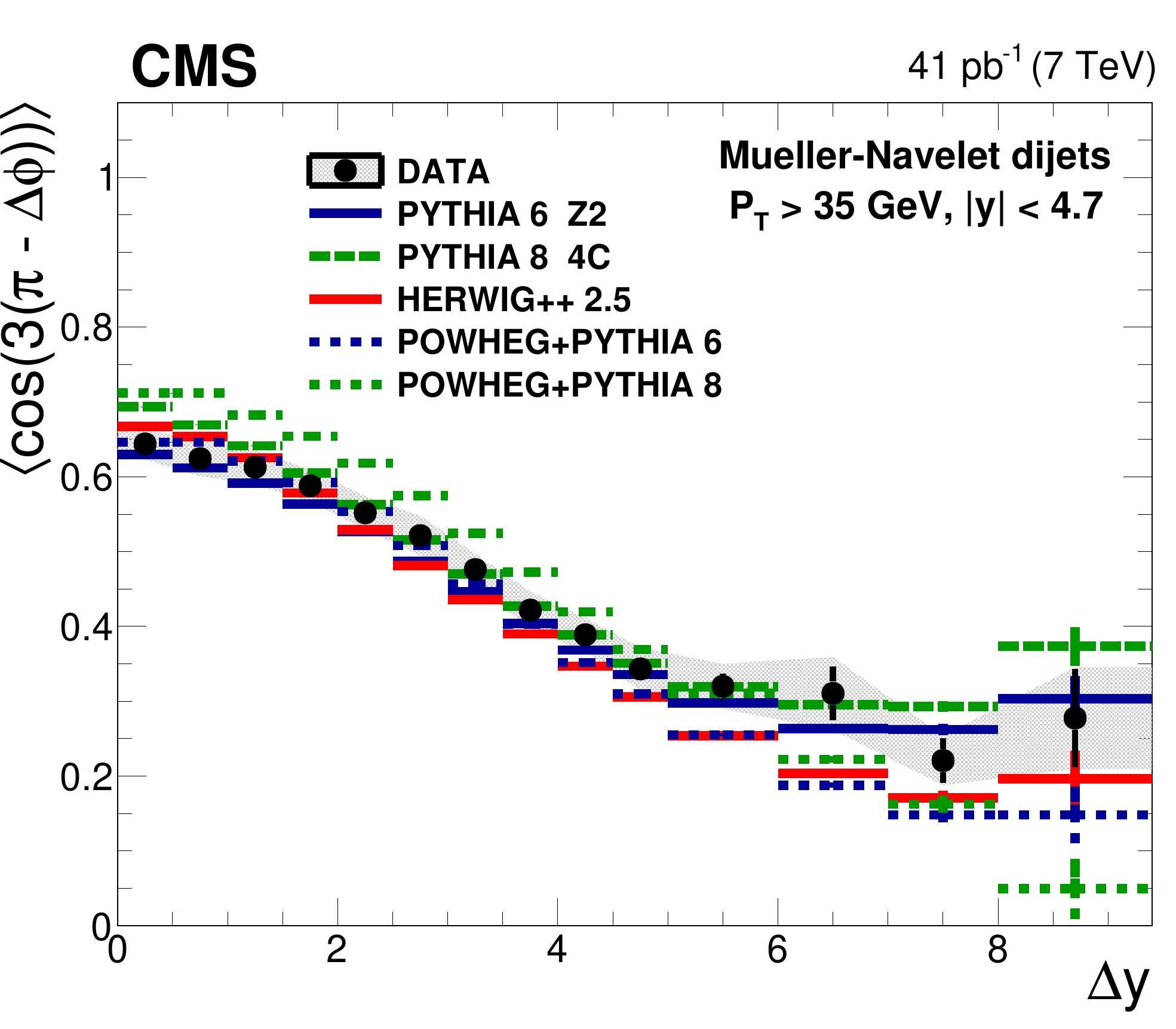}
     \includegraphics[width=0.49\textwidth]{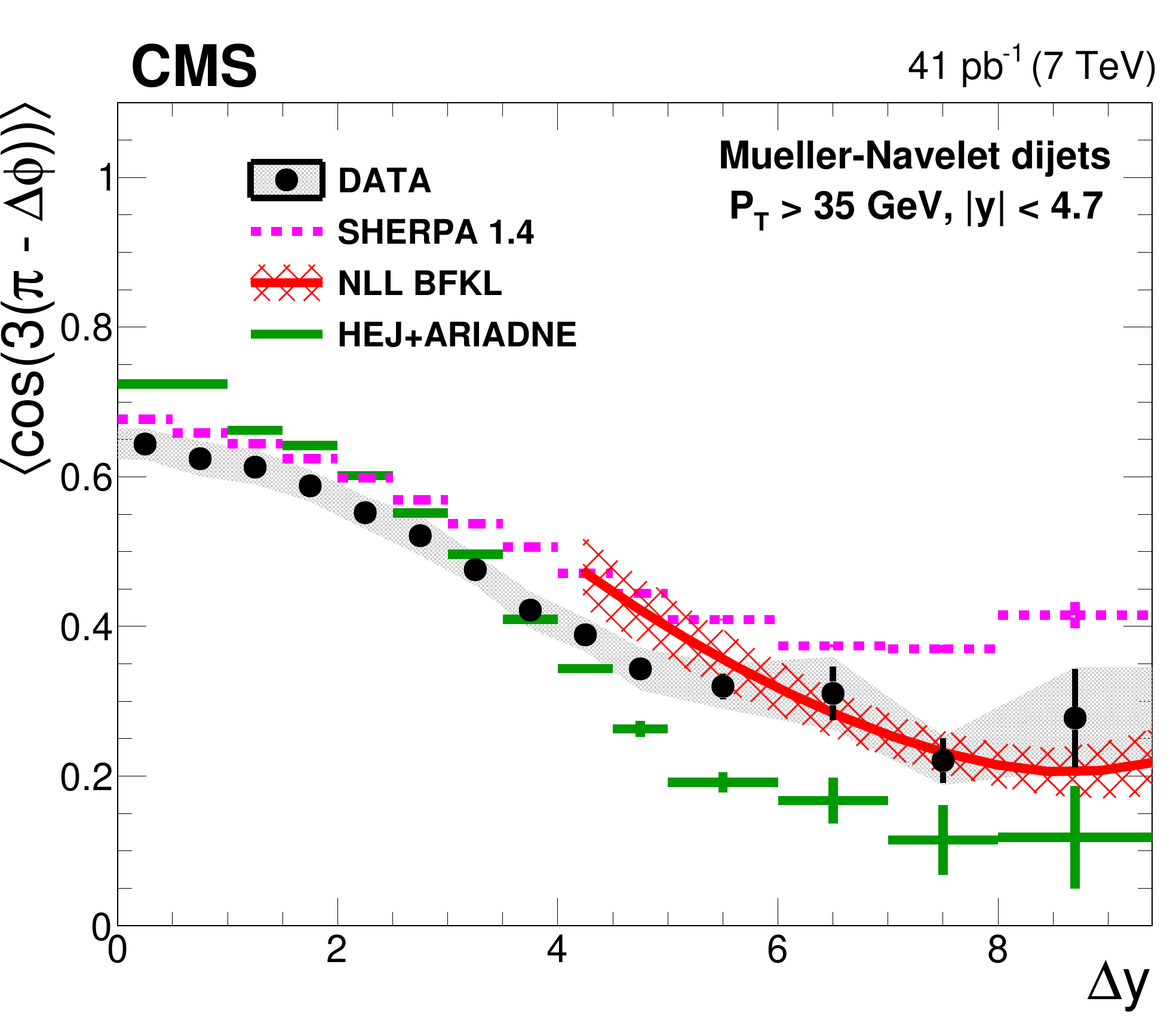}
    \caption{Left: Average $\langle\cos(n(\pi-\Delta\phi)) \rangle$($n = 1,2,3$) as a function of $ \Delta{y} $ compared to LL DGLAP MC generators. In addition, the predictions of the NLO generator \POWHEG interfaced with the LL DGLAP generators \PYTHIA 6 and \PYTHIA 8 are shown. Right: Comparison of the data to the MC generator \SHERPA with parton matrix elements matched to a LL DGLAP parton shower, to the LL BFKL inspired generator  \textsc{hej} with hadronisation by \textsc{ariadne}, and to analytical NLL BFKL calculations at the parton level ($4.0 < \Delta{y} < 9.4$).}
    \label{fig:MN_cos_main}
    \label{fig:MN_cos2_main}
    \label{fig:MN_cos3_main}
  \end{center}
\end{figure}

The measured average cosines, $\langle\cos(n(\pi - \Delta \phi))\rangle$, are less than unity at $\Delta{y} = 0$, due to the emission of jets with $\pt < 35\GeV$. They decrease with increasing $\Delta{y}$, as shown in Fig.~\ref{fig:MN_cos_main}, indicating that the decorrelation of jets increases as the phase space opens up for emission of additional jets with $\pt > 35\GeV$. At large values of the rapidity separation ($\Delta{y} \gtrsim 8$), additional emissions are becoming kinematically suppressed due to energy-momentum conservation near the phase space boundary ($\Delta{y} \approx 10$), resulting in an increase of the average cosines towards unity. In the bin $6 < \Delta{y} < 7$, a flattening of the average cosines is observed. Despite various checks, no systematic effect could be shown to be responsible for this flattening.

In Fig.~\ref{fig:MN_cos_main} (left) the measured average cosines are compared to the predictions obtained from the LL parton shower MC generators  \PYTHIA 6, \HERWIGpp, and \PYTHIA 8. Also shown are the predictions from the NLO \POWHEG generator interfaced with the LL DGLAP generators \PYTHIA 6 and \PYTHIA 8. In Fig.~\ref{fig:MN_cos_main} (right) the measurements are compared to the MC generator \SHERPA, to the \textsc{hej+ariadne} package, and to analytical NLL BFKL calculations at the parton level~\cite{Ducloue:2013bva}.
The comparisons (Fig.~\ref{fig:MN_cos_main}) with the various MC predictions can be summarised as follows: \PYTHIA 6 and \PYTHIA 8 show a slightly stronger decorrelation for the average cosine at large $\Delta{y} $ than observed in the data. For $\langle\cos(2(\pi - \Delta\phi))\rangle$ and $\langle\cos(3(\pi - \Delta\phi))\rangle$ \PYTHIA 6 and \PYTHIA 8 show a fair agreement with the data. \HERWIGpp  shows a satisfactory agreement with the data  on the average cosine. For  $\langle\cos(2(\pi - \Delta\phi))\rangle$ and $\langle\cos(3(\pi - \Delta\phi))\rangle$ \HERWIGpp begins to show a  stronger decorrelation at large $\Delta{y} $ than observed  in the data. The NLO generator \POWHEG interfaced with the two LL DGLAP generators \PYTHIA 6 and \PYTHIA~8 does not improve the agreement with the data obtained with the standalone LL DGLAP generators, while \SHERPA underestimates the azimuthal decorrelation at large $\Delta{y}$ for the measured average cosines. The \textsc{hej+ariadne} package overestimates the azimuthal decorrelation at large $\Delta{y}$ for the measured average cosines.

\begin{figure}[bp]
  \begin{center}
     \includegraphics[width=0.49\textwidth]{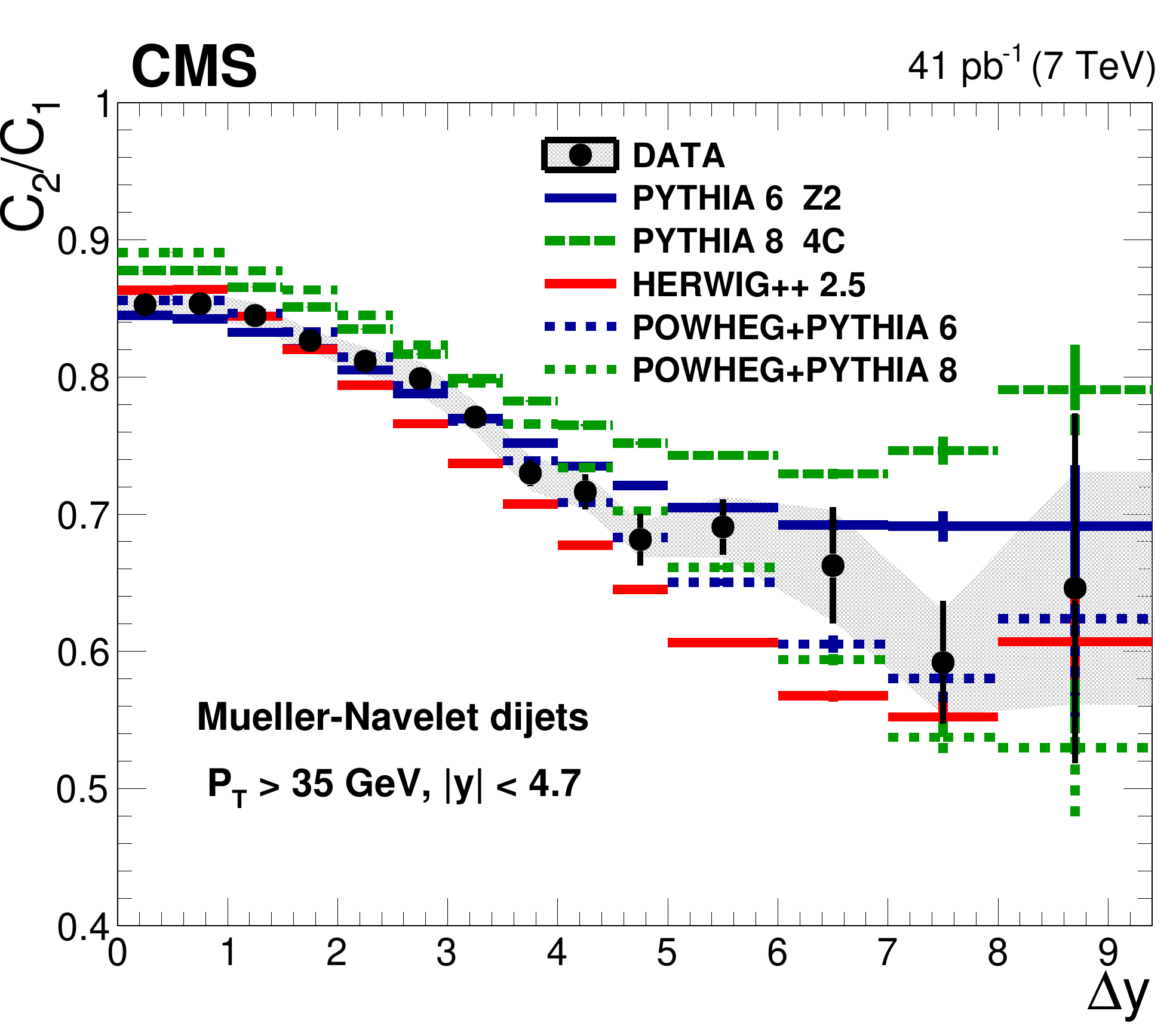}
     \includegraphics[width=0.49\textwidth]{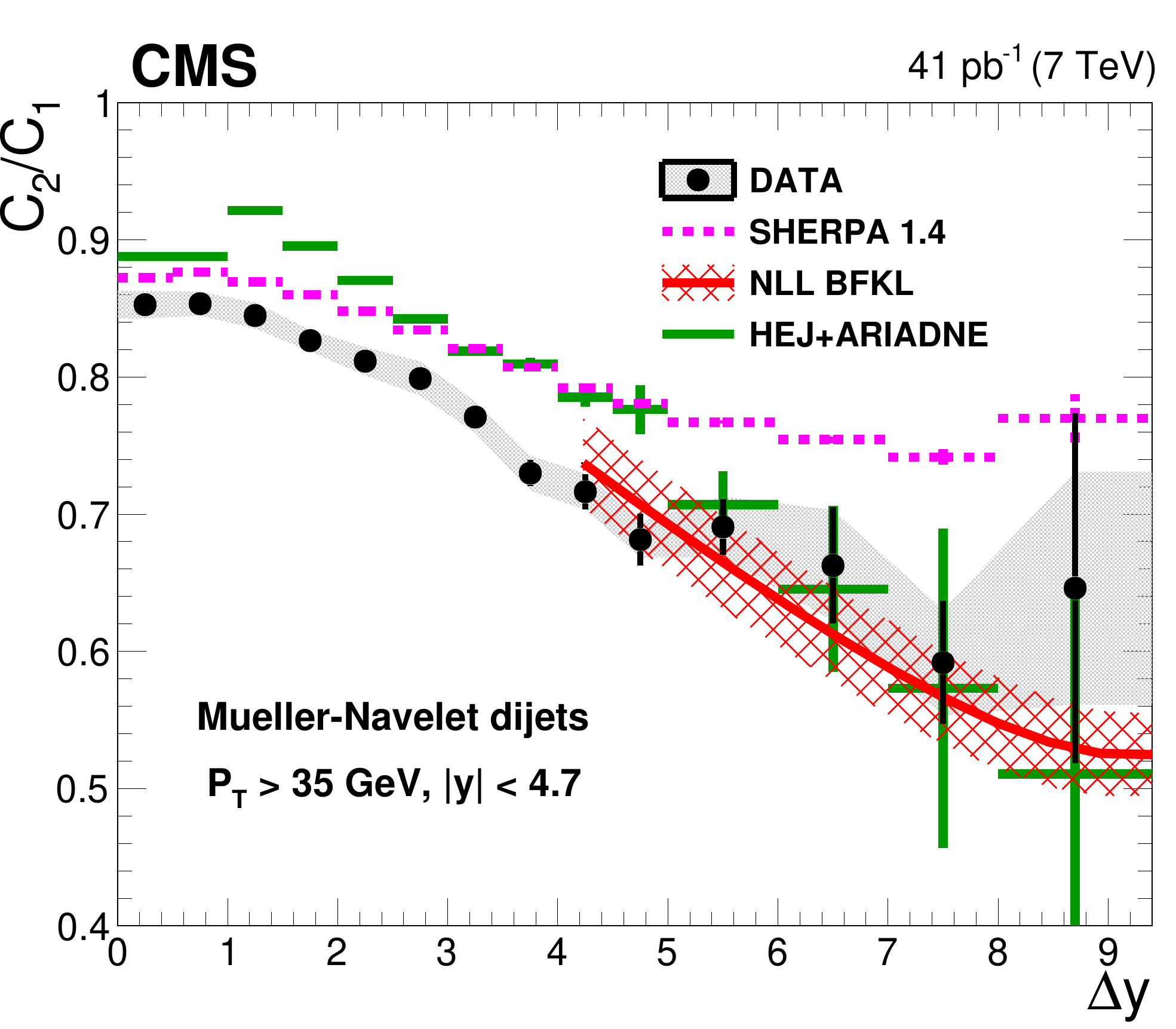}
     \includegraphics[width=0.49\textwidth]{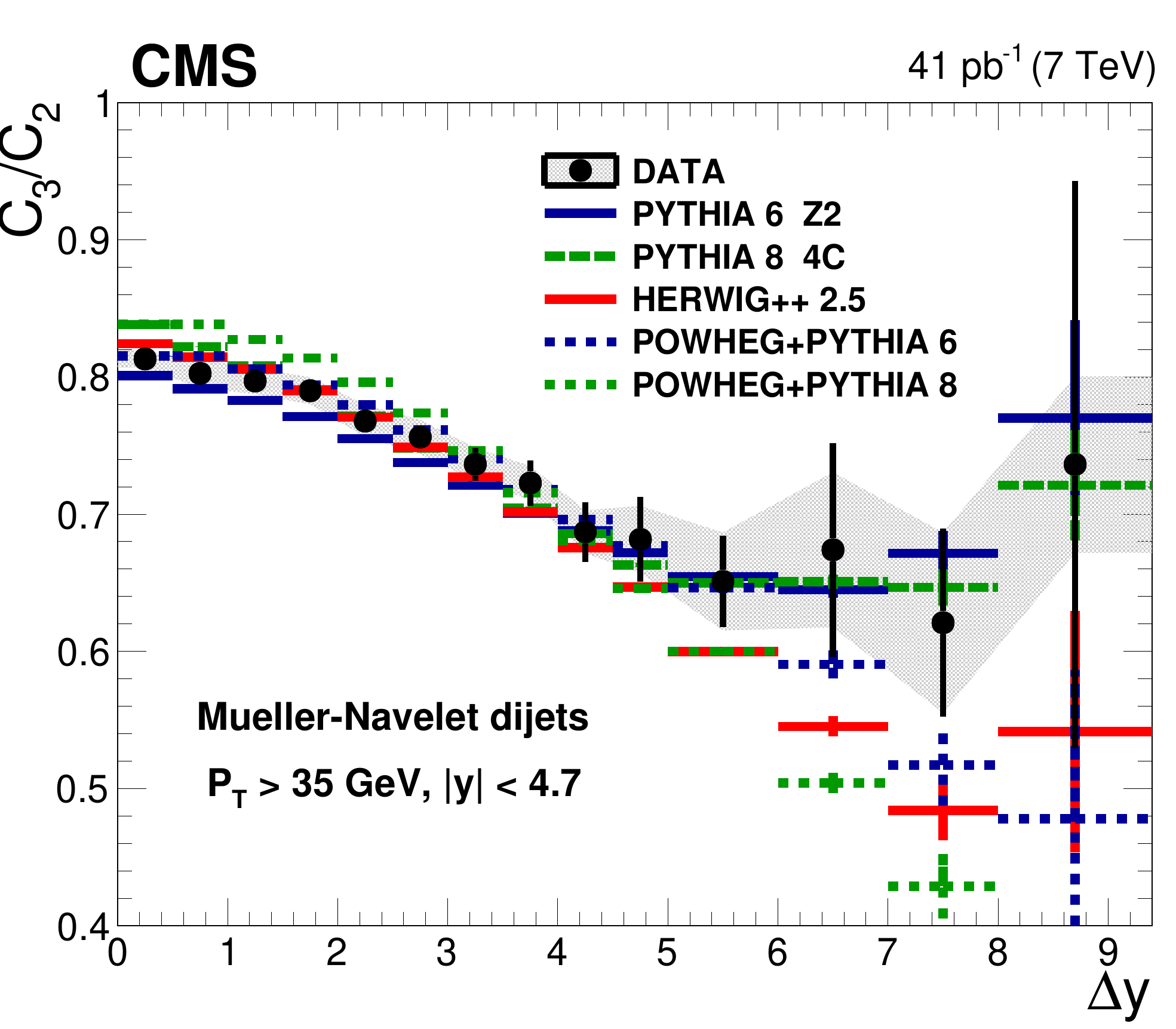}
     \includegraphics[width=0.49\textwidth]{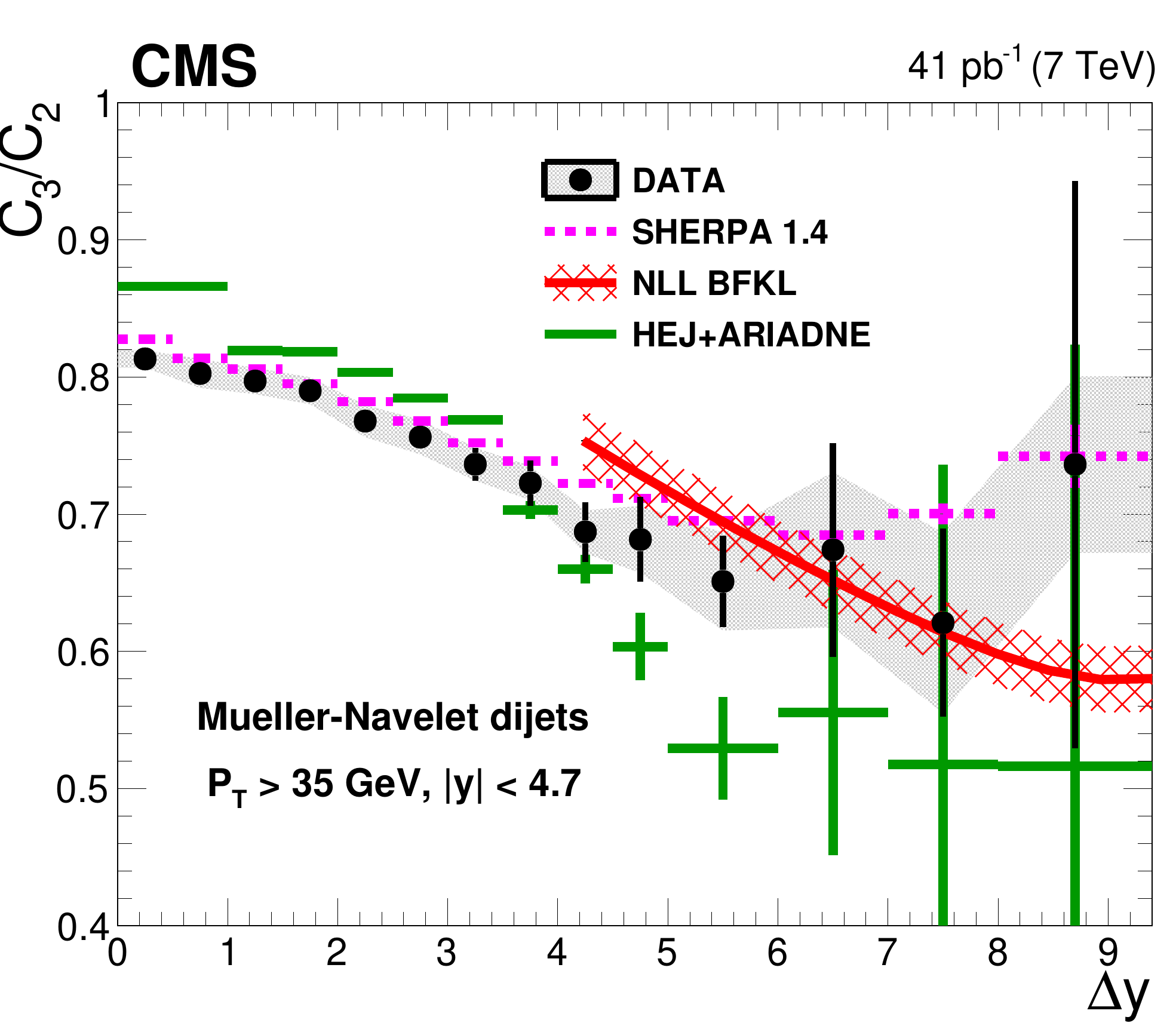}
    \caption{Left: The measured ratios $C_2/C_1$ (top row) and $C_3/C_2$ (bottom row) as a function of rapidity difference $ \Delta{y} $ are compared to LL DGLAP parton shower generators and to the NLO generator \POWHEG interfaced with \PYTHIA 6 and \PYTHIA 8. Right: Comparison of the ratios to the MC generator \SHERPA with parton matrix element matched to a LL DGLAP parton shower, to the LL BFKL-inspired generator  \textsc{hej} with hadronisation by \textsc{ariadne}, and to analytical NLL BFKL calculations at the parton level.}
    \label{fig:MN_rat21_main}
    \label{fig:MN_rat32_main}
  \end{center}
\end{figure}

As mentioned in Section 2, the ratios of cosines are theoretically expected to be more sensitive to BFKL effects than the average cosines and $\Delta\phi$ distributions because of a cancellation of ``pure" DGLAP contributions~\cite{Angioni:2011wj} (although the ratios are still sensitive to the detailed implementation of initial and final state parton radiation, parton showering, and colour coherence effects in each of the MCs).
The measured ratios $C_2/C_1$ and  $C_3/C_2$ as a function of $\Delta{y}$ are shown in Fig.~\ref{fig:MN_rat21_main}. \PYTHIA 6 and \PYTHIA 8 underestimate the azimuthal decorrelation for the average cosine ratio $C_2/C_1$ at large $\Delta{y}$ but are consistent with the data for $C_3/C_2$ within the rather large experimental uncertainties. \HERWIGpp overestimates  the azimuthal decorrelation for the average cosine ratios $C_2/C_1$ and $C_3/C_2$  at large $\Delta{y}$. \SHERPA underestimates the azimuthal decorrelation at large $\Delta{y}$ for the average cosine ratio $C_2/C_1$ but is consistent with the data for $C_3/C_2$ within the experimental uncertainties. The \textsc{hej+ariadne} package overestimates the azimuthal decorrelation at large $\Delta{y}$ for the average cosine ratios $C_2/C_1$ and $C_3/C_2$.

The analytical NLL BFKL calculations performed at the parton level~\cite{Ducloue:2013bva} agree well with the data for all measured observables within the experimental and theoretical uncertainties. The predictions are based on a full NLL BFKL calculation~\cite{Ducloue:2013hia,Caporale:2012ih}, which is improved by a generalised optimal choice of the renormalisation scale~\cite {Brodsky:1982gc, Brodsky:1998kn}, and available for the $ \Delta{y}$ range from $4.0$ to $9.4$.

The uncertainties on the NLL BFKL predictions in Fig.~\ref{fig:MN_dphi_0-3} (bottom row) and Fig.~\ref{fig:MN_cos_main} (right) are obtained by variation of the parameters of the NLL BFKL approximation (renormalisation and factorisation scales). Thus, theoretical uncertainties on the NLL BFKL predictions in Fig.~\ref{fig:MN_rat21_main} (right) consist just of those due to missing higher-order corrections. The NLL BFKL calculation performed by a different group of authors showed worse agreement with these data~\cite{Caporale:2014gpa}.

The measured data are also compared to predictions of the LL BFKL-motivated MC generator \textsc{cascade} 2~\cite{Jung:2010si} (not shown), which is based on the CCFM evolution equation~\cite{Jung:2000hk}, and which shows an even stronger decorrelation than that predicted by the \textsc{hej+ariadne} package.

\begin{figure}[hbtp]
  \begin{center}
     \includegraphics[width=0.49\textwidth]{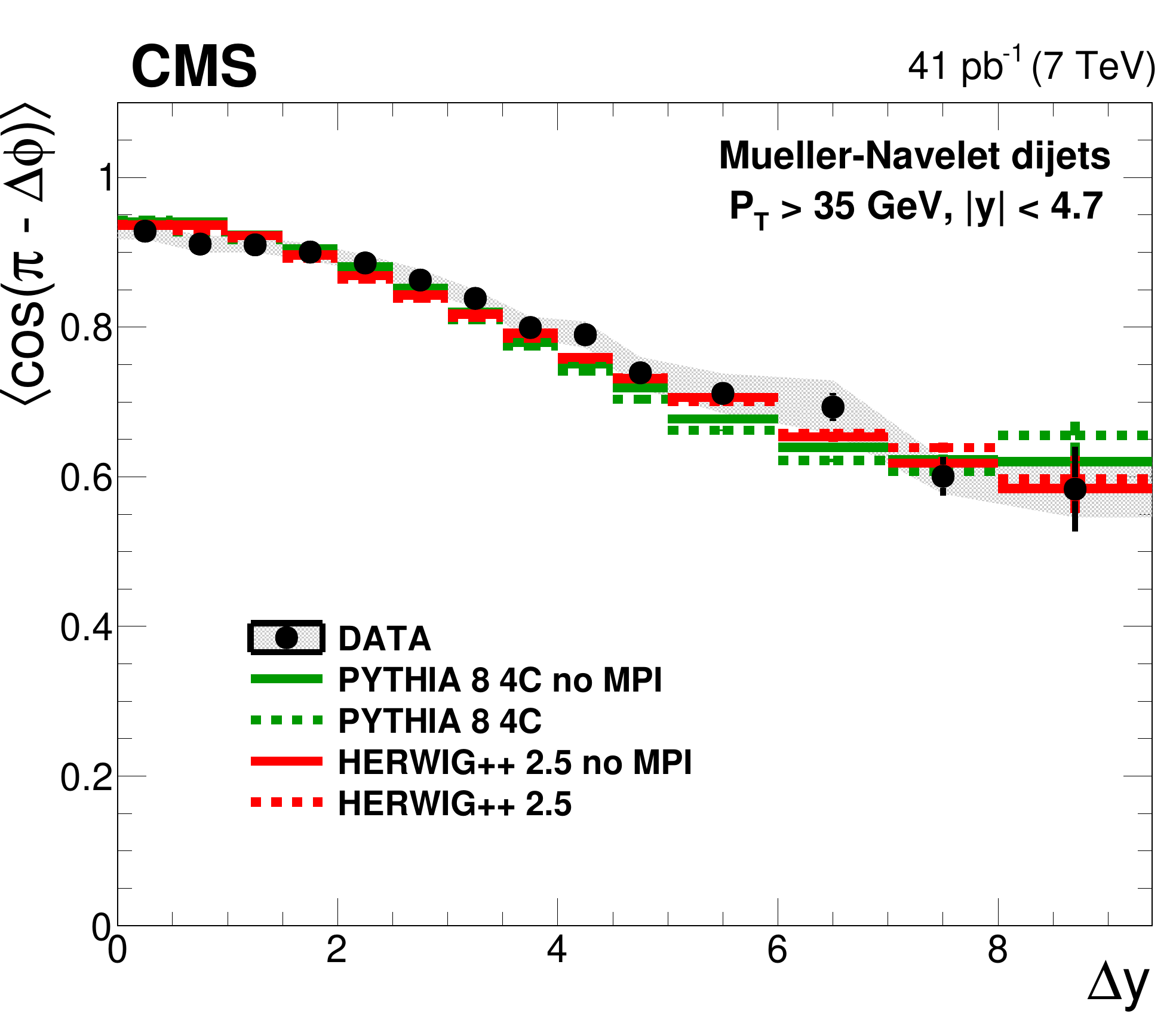}
     \includegraphics[width=0.49\textwidth]{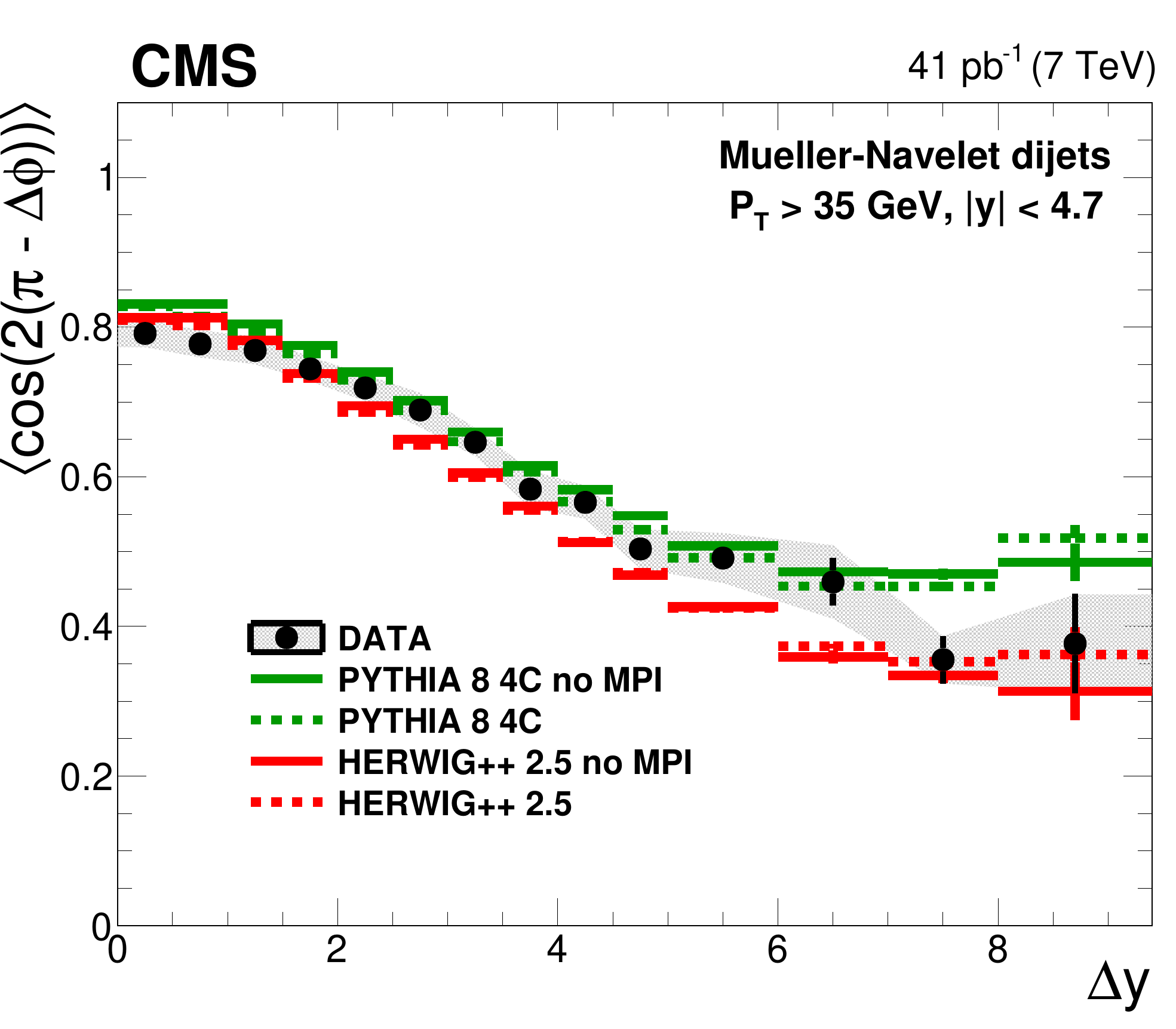}
     \includegraphics[width=0.49\textwidth]{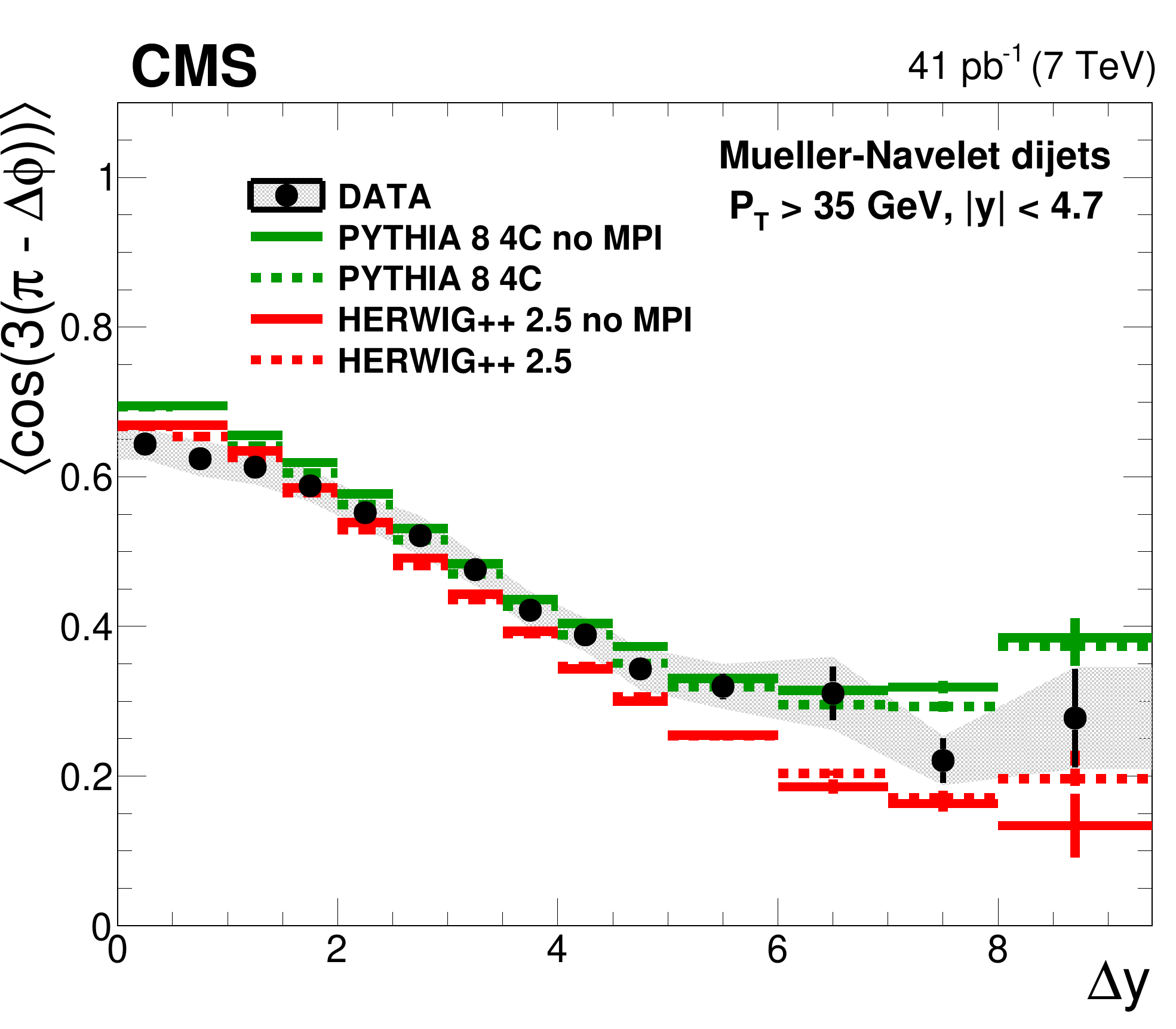}
    \caption{Average $\langle\cos{(\pi - \Delta \phi)}\rangle$, $\langle\cos{2(\pi - \Delta \phi)}\rangle$ and $\langle\cos{3(\pi - \Delta \phi)}\rangle$ compared to \PYTHIA 6 with and without MPI.}
    \label{fig:MN_cos_AOMPI_Pythia8}
  \end{center}
\end{figure}

\begin{figure}[hbtp]
  \begin{center}
     \includegraphics[width=0.49\textwidth]{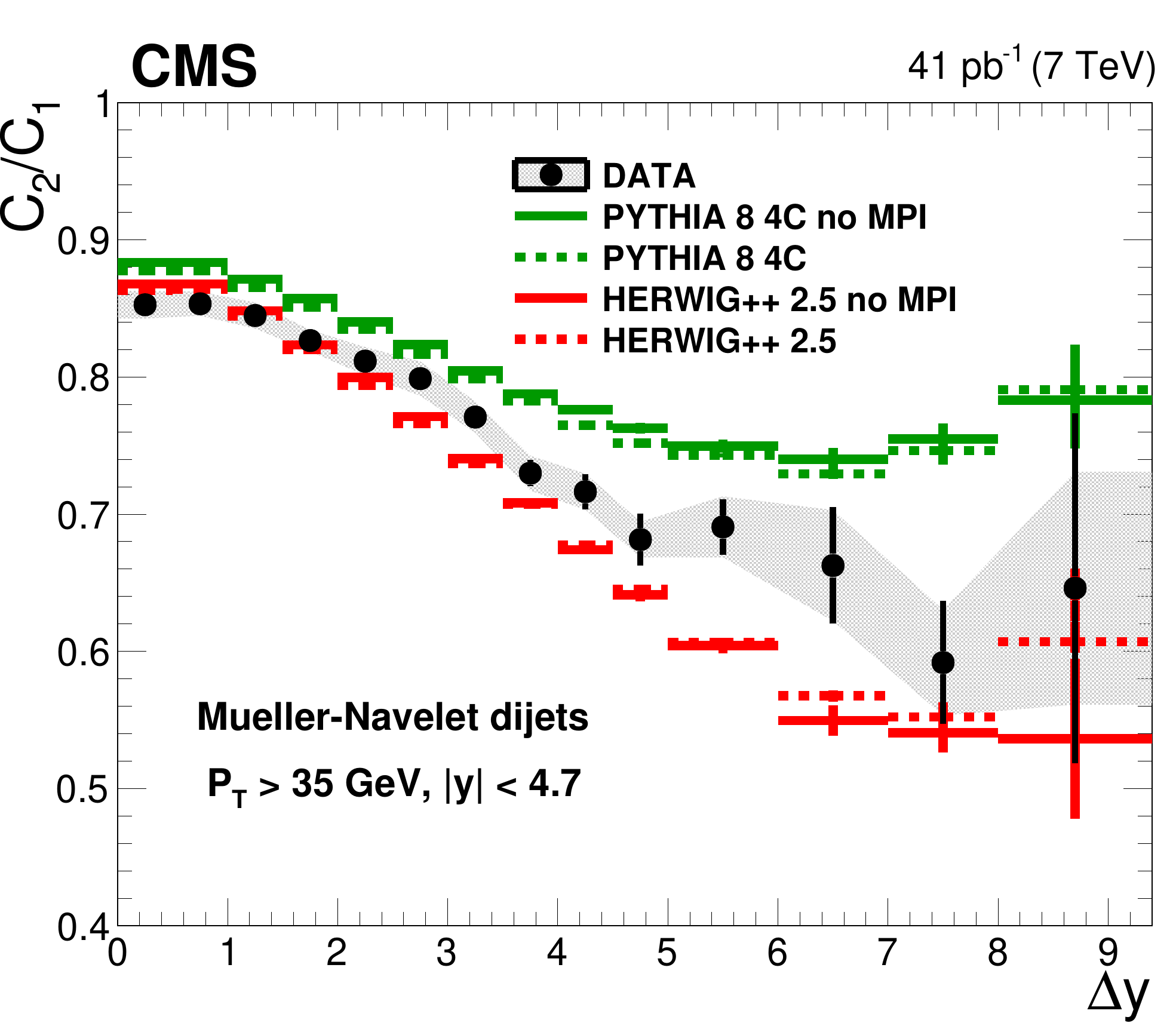}
     \includegraphics[width=0.49\textwidth]{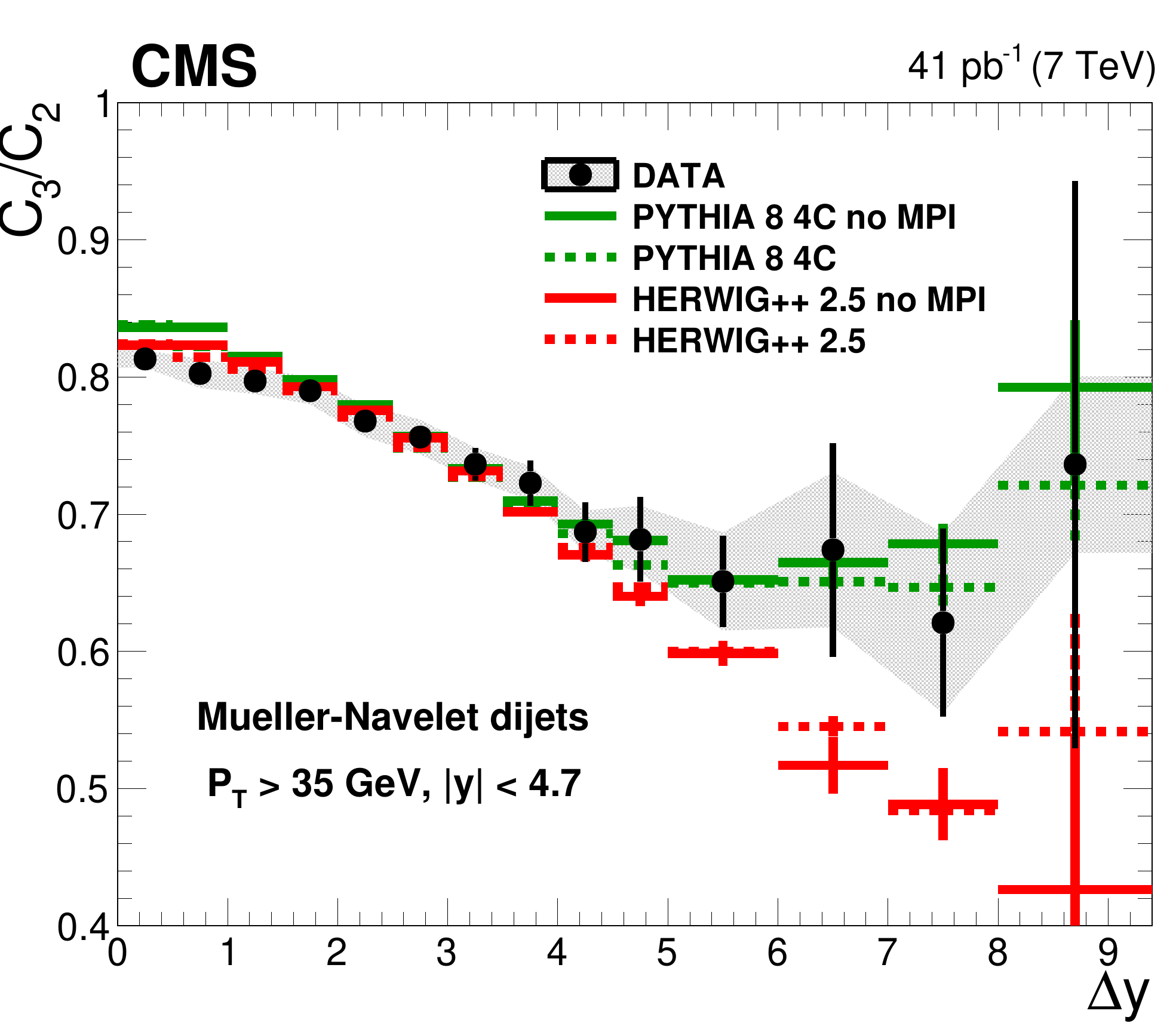}
    \caption{Measured ratios $C_2/C_1$ (left) and $C_3/C_2$ (right) compared to \PYTHIA 8 with and without MPI.}
    \label{fig:MN_rats_AOMPI_Pythia8}
  \end{center}
\end{figure}

Multiparton interactions (MPI) are an additional source of azimuthal decorrelation since they can produce additional jets not correlated with those from the primary interaction. By default, MPI effects are included in the MC generators \PYTHIA 6, \PYTHIA 8 , \HERWIGpp, and \SHERPA. In order to study the influence of the MPI on the azimuthal decorrelation, the corresponding options in the MC generators are used to disable the MPI modelling. The measurements are then compared with the \PYTHIA 8 and \HERWIGpp predictions with and without MPI in Figs.~\ref{fig:MN_cos_AOMPI_Pythia8} and \ref{fig:MN_rats_AOMPI_Pythia8}, where it can be seen that the average cosines are not sensitive to the details of MPI modelling in \PYTHIA 8 and \HERWIGpp. Other generators show an even smaller spread of predictions with and without MPI.

Another potential source of azimuthal decorrelation is the hadronisation of the produced partons, which can potentially smear out their azimuthal angle. The size of this non-perturbative effect is estimated by a comparison of observables at the parton and stable-particle levels,
as obtained with \PYTHIA 6. The observed variations in the measured observables do not exceed 10\%.
It is found that, in general, the size of hadronisation and MPI effects does not significantly exceed the experimental uncertainties, justifying a  direct comparison of the analytical NLL BFKL calculations~\cite{Ducloue:2013bva} performed at the parton level with the measured observables.

It should be noted that all DGLAP MC generators used in this work incorporate colour-coherence effects (colour dipoles, polar-angle ordering, etc.), which are rapidity-dependent parton radiation effects that complement the DGLAP evolution. Taking these effects into account at small $\Delta{y}$, where $(\alpha_S \, \Delta{y})^n $ terms are small (i.e. in the DGLAP domain), leads to an improvement of data description, while at large $\Delta{y}$ they yield a worse description of the data. As a matter of fact, different implementations of colour-coherence effects in the DGLAP MC generators result in similar effects at small $\Delta{y} $, but in quite different predictions for the large $\Delta{y}$ region for dijet ratios~\cite{Chatrchyan:2012pb} and for the azimuthal decorrelation observables presented here. A better theoretical prediction might be obtained if these $\Delta{y}$ dependent contributions are replaced by the complete BFKL calculation at large $\Delta{y}$, where $(\alpha_S \, \Delta{y})^n $ terms are large and the BFKL approach is expected to be more reliable.

\section{Conclusions}

The first measurement of the azimuthal decorrelation of the most-forward and backward jets in the event (called Mueller--Navelet dijets), with rapidity separations up to $\Delta{y}=9.4$, is presented for proton-proton collisions at $\sqrt{s}=7\TeV$. The measured observables include azimuthal-angle distributions, moments of the average cosines of the decorrelation angle, $\langle\cos(n(\pi - \Delta \phi))\rangle$ for $n=1$, $2$, $3$, as well as ratios of the average cosines, as a function of the rapidity separation $\Delta{y}$ between the MN jets.

The predictions of the DGLAP-based MC generator \HERWIGpp 2.5, improved with leading-log (LL) parton showers and colour-coherence effects, exhibit satisfactory agreement with the data for all measured observables. Other MC generators of this type, such as \PYTHIA 6 Z2, \PYTHIA 8 4C, and \SHERPA 1.4, provide a less accurate description of all measurements.

The MC generator \POWHEG, with NLO matrix elements interfaced with the LL parton shower of \PYTHIA~6 and \PYTHIA~8, does not improve the overall agreement with the data compared to the description provided by \PYTHIA~6 and 8 alone.

The MC generator \textsc{hej}, based on LL BFKL matrix elements combined with \textsc{ariadne} for parton shower and hadronisation, predicts a stronger decorrelation than observed in the data.

An analytical BFKL calculation at next-to-leading logarithmic (NLL) accuracy with an optimised renormalisation scheme and scale, provides a satisfactory description of the data for the measured jet observables at $\Delta{y}>4$.

The observed sensitivity to the implementation of the colour-coherence effects in the DGLAP MC generators and the reasonable data-theory agreement shown by the NLL BFKL analytical calculations at large  $\Delta{y}$, may be considered as indications that the kinematical domain of the present study lies in between the regions described by the DGLAP and BFKL approaches. Possible manifestations of BFKL signatures are expected to be more pronounced at increasing collision energies.

\section*{Acknowledgements}

We would like to thank Bertrand Duclou{\'e}, Lech Szymanowski, and Samuel Wallon for providing the full NLL BFKL analytical predictions. We also thank Lev Lipatov for useful discussions.

\hyphenation{Bundes-ministerium Forschungs-gemeinschaft Forschungs-zentren} We congratulate our colleagues in the CERN accelerator departments for the excellent performance of the LHC and thank the technical and administrative staffs at CERN and at other CMS institutes for their contributions to the success of the CMS effort. In addition, we gratefully acknowledge the computing centres and personnel of the Worldwide LHC Computing Grid for delivering so effectively the computing infrastructure essential to our analyses. Finally, we acknowledge the enduring support for the construction and operation of the LHC and the CMS detector provided by the following funding agencies: the Austrian Federal Ministry of Science, Research and Economy and the Austrian Science Fund; the Belgian Fonds de la Recherche Scientifique, and Fonds voor Wetenschappelijk Onderzoek; the Brazilian Funding Agencies (CNPq, CAPES, FAPERJ, and FAPESP); the Bulgarian Ministry of Education and Science; CERN; the Chinese Academy of Sciences, Ministry of Science and Technology, and National Natural Science Foundation of China; the Colombian Funding Agency (COLCIENCIAS); the Croatian Ministry of Science, Education and Sport, and the Croatian Science Foundation; the Research Promotion Foundation, Cyprus; the Ministry of Education and Research, Estonian Research Council via IUT23-4 and IUT23-6 and European Regional Development Fund, Estonia; the Academy of Finland, Finnish Ministry of Education and Culture, and Helsinki Institute of Physics; the Institut National de Physique Nucl\'eaire et de Physique des Particules~/~CNRS, and Commissariat \`a l'\'Energie Atomique et aux \'Energies Alternatives~/~CEA, France; the Bundesministerium f\"ur Bildung und Forschung, Deutsche Forschungsgemeinschaft, and Helmholtz-Gemeinschaft Deutscher Forschungszentren, Germany; the General Secretariat for Research and Technology, Greece; the National Scientific Research Foundation, and National Innovation Office, Hungary; the Department of Atomic Energy and the Department of Science and Technology, India; the Institute for Studies in Theoretical Physics and Mathematics, Iran; the Science Foundation, Ireland; the Istituto Nazionale di Fisica Nucleare, Italy; the Ministry of Science, ICT and Future Planning, and National Research Foundation (NRF), Republic of Korea; the Lithuanian Academy of Sciences; the Ministry of Education, and University of Malaya (Malaysia); the Mexican Funding Agencies (CINVESTAV, CONACYT, SEP, and UASLP-FAI); the Ministry of Business, Innovation and Employment, New Zealand; the Pakistan Atomic Energy Commission; the Ministry of Science and Higher Education and the National Science Centre, Poland; the Funda\c{c}\~ao para a Ci\^encia e a Tecnologia, Portugal; JINR, Dubna; the Ministry of Education and Science of the Russian Federation, the Federal Agency of Atomic Energy of the Russian Federation, Russian Academy of Sciences, and the Russian Foundation for Basic Research; the Ministry of Education, Science and Technological Development of Serbia; the Secretar\'{\i}a de Estado de Investigaci\'on, Desarrollo e Innovaci\'on and Programa Consolider-Ingenio 2010, Spain; the Swiss Funding Agencies (ETH Board, ETH Zurich, PSI, SNF, UniZH, Canton Zurich, and SER); the Ministry of Science and Technology, Taipei; the Thailand Center of Excellence in Physics, the Institute for the Promotion of Teaching Science and Technology of Thailand, Special Task Force for Activating Research and the National Science and Technology Development Agency of Thailand; the Scientific and Technical Research Council of Turkey, and Turkish Atomic Energy Authority; the National Academy of Sciences of Ukraine, and State Fund for Fundamental Researches, Ukraine; the Science and Technology Facilities Council, UK; the US Department of Energy, and the US National Science Foundation.

Individuals have received support from the Marie-Curie programme and the European Research Council and EPLANET (European Union); the Leventis Foundation; the A. P. Sloan Foundation; the Alexander von Humboldt Foundation; the Belgian Federal Science Policy Office; the Fonds pour la Formation \`a la Recherche dans l'Industrie et dans l'Agriculture (FRIA-Belgium); the Agentschap voor Innovatie door Wetenschap en Technologie (IWT-Belgium); the Ministry of Education, Youth and Sports (MEYS) of the Czech Republic; the Council of Science and Industrial Research, India; the HOMING PLUS programme of the Foundation for Polish Science, cofinanced from European Union, Regional Development Fund; the OPUS programme of the National Science Center (Poland); the Compagnia di San Paolo (Torino); MIUR project 20108T4XTM (Italy); the Thalis and Aristeia programmes cofinanced by EU-ESF and the Greek NSRF; the National Priorities Research Program by Qatar National Research Fund; the Rachadapisek Sompot Fund for Postdoctoral Fellowship, Chulalongkorn University (Thailand); the Chulalongkorn Academic into Its 2nd Century Project Advancement Project (Thailand); and the Welch Foundation, contract C-1845.

\bibliography{auto_generated}

\cleardoublepage \appendix\section{The CMS Collaboration \label{app:collab}}\begin{sloppypar}\hyphenpenalty=5000\widowpenalty=500\clubpenalty=5000\textbf{Yerevan Physics Institute,  Yerevan,  Armenia}\\*[0pt]
V.~Khachatryan, A.M.~Sirunyan, A.~Tumasyan
\vskip\cmsinstskip
\textbf{Institut f\"{u}r Hochenergiephysik der OeAW,  Wien,  Austria}\\*[0pt]
W.~Adam, E.~Asilar, T.~Bergauer, J.~Brandstetter, E.~Brondolin, M.~Dragicevic, J.~Er\"{o}, M.~Flechl, M.~Friedl, R.~Fr\"{u}hwirth\cmsAuthorMark{1}, V.M.~Ghete, C.~Hartl, N.~H\"{o}rmann, J.~Hrubec, M.~Jeitler\cmsAuthorMark{1}, V.~Kn\"{u}nz, A.~K\"{o}nig, M.~Krammer\cmsAuthorMark{1}, I.~Kr\"{a}tschmer, D.~Liko, T.~Matsushita, I.~Mikulec, D.~Rabady\cmsAuthorMark{2}, B.~Rahbaran, H.~Rohringer, J.~Schieck\cmsAuthorMark{1}, R.~Sch\"{o}fbeck, J.~Strauss, W.~Treberer-Treberspurg, W.~Waltenberger, C.-E.~Wulz\cmsAuthorMark{1}
% ...
\vskip\cmsinstskip
% ...
1:~~Also at Vienna University of Technology, Vienna, Austria\\
2:~~Also at CERN, European Organization for Nuclear Research, Geneva, Switzerland\\
% ...

\end{sloppypar}
\end{document}